\documentclass[12pt,showpacs,preprintnumbers,floatfix]{article}
\usepackage{graphicx} 
\usepackage{dcolumn} 
\usepackage{bm} 
\usepackage{amsfonts}
\usepackage{amsthm}
\usepackage{amsmath}
\usepackage{amssymb}
\usepackage{epsfig}
\usepackage{color}
\usepackage{textcomp}
\usepackage{hyperref}
\usepackage{titlesec}
\usepackage{slashed}
\usepackage{caption}
\usepackage{subcaption}
\usepackage{siunitx}
\usepackage{booktabs}
\usepackage{multirow}
\usepackage{cite}
\usepackage{placeins}

\def\be{\begin{equation}}
\def\ee{\end{equation}}
\def\ba{\begin{eqnarray}}
\def\ea{\end{eqnarray}}

\def\bdm{\begin{displaymath}}
\def\edm{\end{displaymath}}

\def\bq{\begin{quote}}
\def\eq{\end{quote}}

 at 10truept

\newcommand{\bea}{\begin{eqnarray}}
\newcommand{\eea}{\end{eqnarray}}

\newcommand{\bi}{\begin{itemize}}
\newcommand{\ei}{\end{itemize}}

\newcommand{\beq}{\begin{equation}}
\newcommand{\eeq}{\end{equation}}
\newcommand{\beqa}{\begin{eqnarray}}
\newcommand{\eeqa}{\end{eqnarray}}

\newcommand{\nn}{\nonumber}

\def\ltap{\ \raise.3ex\hbox{$<$\kern-.75em\lower1ex\hbox{$\sim$}}\ }
\def\gtap{\ \raise.3ex\hbox{$>$\kern-.75em\lower1ex\hbox{$\sim$}}\ }
\def\gl{\ \raise.5ex\hbox{$>$}\kern-.8em\lower.5ex\hbox{$<$}\ }
\def\roughly#1{\raise.3ex\hbox{$#1$\kern-.75em\lower1ex\hbox{$\sim$}}}

\begin{document}

\begin{flushright}
{\small
TUM-HEP-1167/18\\
\today
}
\end{flushright}

\vspace*{2cm}

\begin{center}
{\Large \bf Charged Planckian Interacting Dark Matter}\\

\vspace*{1.0cm} 
{\large Mathias Garny$^{a}$\footnote{\tt mathias.garny@tum.de}, 
Andrea Palessandro$^{b}$\footnote{\tt palessandro@cp3.sdu.dk},\\[0.5ex] 
McCullen Sandora$^{c}$\footnote{\tt mccullen.sandora@tufts.edu},  
Martin S. Sloth$^{b}$\footnote{\tt sloth@cp3.sdu.dk}}\\
\vspace{.5cm} {\em $^a$Physik Department T31, Technische Universit\"at M\"unchen,\\
James-Franck-Stra\ss e 1, D-85748 Garching, Germany}\\
\vspace{.5cm} {\em $^b$CP$^3$-Origins, Center for Cosmology and Particle Physics Phenomenology \\ University of Southern Denmark, Campusvej 55, 5230 Odense M, Denmark}\\
\vspace{.5cm} {\em  $^c$Institute of Cosmology, Department of Physics and Astronomy\\ Tufts University, Medford, MA 02155, USA}

\end{center}

\begin{abstract}
A minimal model of Cold Dark Matter (CDM) is a very massive particle with only gravitational interactions, also called Planckian Interacting Dark Matter (PIDM). Here we consider an extension of the PIDM framework by an unbroken $U(1)$ gauge symmetry under which the PIDM is charged, but remains only gravitationally coupled to the Standard Model (SM). Contrary to ``hidden charged dark matter", the charged PIDM never reaches thermal equilibrium with the SM. 
The dark sector is populated by freeze-in via gravitational interactions at reheating. If the dark fine-structure constant $\alpha_D$ is larger than about $10^{-3}$,
the dark sector thermalizes within itself, and the PIDM abundance is further modified by freeze-out in the dark sector. Interestingly, this largely reduces
the dependence of the final abundance on the reheating temperature, as compared to an uncharged PIDM.
Thermalization within the dark sector is driven by inelastic radiative processes, and affected by the Landau-Pomeranchuk-Migdal (LPM) effect.
The observed CDM abundance can be obtained over a wide mass range from the weak to the GUT scale, and for phenomenologically interesting
couplings $\alpha_D\sim 10^{-2}$. 
Due to the different thermal history, the charged PIDM can be discriminated from ``hidden charged dark matter" by 
more precise measurements of the effective number of neutrino species $N_{\rm eff}$.

\end{abstract}

\newpage

\section{Introduction}

The current standard model of cosmology, the $\Lambda$CDM model, is favoured as the simplest model that provides a good fit
to a wide range of cosmological observations to date \cite{Aghanim:2018eyx, Anderson:2013zyy, Riess:2016jrr}. This implies that dark matter is well described by non-relativistic particles that interact only gravitationally, and, arguably, the minimal model of dark matter from an Occam's razor point of view is therefore Planckian Interacting Dark Matter (PIDM) \cite{Garny:2015sjg,Garny:2017kha} (for earlier related work see \cite{Kolb:1998ki,Chung:2001cb}, and for related subsequent work see \cite{PIDMrel}). 

Although there are good theoretical reasons to consider more elaborate forms of dark matter, such as WIMPs or axions, these models have so far escaped detection. The lack of confirmation of WIMPs in particular has stimulated the scientific community to think more generally about dark matter: one such model is hidden charged dark matter, where dark matter is charged under its own dark force that does not directly couple to the Standard Model (SM) of particle physics \cite{Foot:2004pa,Feng:2008mu,Ackerman:mha,Feng:2009mn,Feng:2009hw,Das:2010ts,Agrawal:2016quu,Tulin:2012wi}. One of the motivations for this type of self-interacting dark matter have been discrepancies between numerical simulations of structure formation in the framework of collisionless cold dark matter (CDM) and observations on galactic and sub-galactic scales \cite{Tulin:2017ara}. 
While these so-called small scale problems of $\Lambda$CDM could well be resolved by a better understanding of the complex baryonic and astrophysical processes relevant on these scales \cite{Bullock:2017xww}, one may ask whether deviations from the collisionless CDM paradigm can be probed by taking advantage of the large amount of observational
data related to the dynamics and kinematics at galactic scales, when adopting a conservative attitude towards the treatment of uncertainties. 

In particular, it has been argued that charged dark matter is already too constrained, for example by galactic triaxiality \cite{Feng:2009mn} and disruption of
dwarf galaxies passing through the host halo \cite{Kahlhoefer:2013dca}, to provide an explanation of the aforementioned small scale problems of $\Lambda$CDM. 
However, this finding has recently been challenged by Agrawal, Cyr-Racine, Randall and Scholtz \cite{Agrawal:2016quu}, who argue that present astrophysical uncertainties
do not allow one to firmly draw this conclusion. 
Although it is also very possible that the small scale problems might have more conventional explanations relying on baryonic physics, it is a simple and phenomenologically interesting possibility that dark matter could be charged under a dark force.

If dark matter is truly hidden, with interactions with the SM suppressed by the GUT scale or higher, then, with the most recent constraint on the scale of inflation \cite{Akrami:2018odb}, dark matter can never have been in thermal equilibrium with the SM (as we discuss in section \ref{xiO1}). In this case dark matter could instead have been produced out of equilibrium in a ``freeze-in" process due to non-renormalizable interactions, where the production is dominated by the highest available temperatures \cite{Chung:1998ua,Chung:1998rq,Hall:2009bx}. This scenario therefore requires a large reheating temperature, implying a lower bound on the amplitude of
primordial gravitational waves.

Previously, in the discussion of hidden charged dark matter, it has typically been assumed in the literature that dark matter as well as the gauge bosons belonging to the new gauge interaction under which dark matter is charged (dark photons) are initially in thermal equilibrium, with an initial temperature similar to the temperature of SM radiation. In those models dark matter is a thermal relic and the dark matter abundance is given by freeze-out in the dark sector \cite{Foot:2004pa,Feng:2008mu,Ackerman:mha,Feng:2009mn,Feng:2009hw,Das:2010ts,Agrawal:2016quu,Tulin:2012wi}. This assumes  that dark matter must have been in thermal equilibrium with the SM early on, and so dark matter can not be maximally hidden in these models, and certainly it cannot be the PIDM. 

Our goal here is therefore to investigate the abundance calculation and the phenomenology in the case of a maximally hidden charged PIDM. We have already previously computed the dark matter abundance for an uncharged PIDM~\cite{Garny:2015sjg, Garny:2017kha}. The additional interactions related to the charge under the dark force,
and the presence of the dark force carrier, can potentially change the freeze-in production as well as the subsequent evolution in the dark sector. As we will see, this affects the final dark matter abundance, and has an impact on the effective number of relativistic degrees of freedom.  We show that the charged PIDM  can be produced with the required abundance to account for the observed dark matter density for ${\cal O}(100)$\,GeV mass of dark matter and self-interactions that are strong enough to potentially have an impact on the small scale problems of CDM. In the case that these ``problems'' are explained by standard baryonic physics, we find that self-interactions are a simple generalization of the PIDM framework, that can be tested in the small mass region of parameter space.

One way to state the philosophy behind the present work is the following: traditional naturalness considerations are so far betraying us as a guide to new physics, and so far Occam's razor in its most naive form has been a better guide for understanding dark energy, the SM Higgs sector and dark matter. Following Occam's razor we may therefore take the simplest ansatz for dark matter as just a massive particle with only gravitational interactions, which can still fit the data. However, if the small scale problems of CDM are indications of self-interactions in the dark sector, it may again be minimally accommodated by extending the model with only one new free parameter, $\alpha_D$, by adding an unbroken $U(1)$ gauge symmetry. In any case it is important to constrain this simple scenario as much as possible, and understand exactly how much we can learn about dark matter in this approach.

An outline of the paper is as follows: In section \ref{sec:setup} we discuss the theoretical setup and define the charged PIDM model. In section \ref{pheno} we present the phenomenology of the model, including a discussion of the dark matter abundance and the
thermal history depending on the value of the dark fine-structure constant $\alpha_D$, focusing mostly on the regime in which the charged PIDM can affect small scale features of dark matter. Some technical details related to non-thermal distribution functions are delegated to the appendix.
We also briefly comment on the regime of a very heavy PIDM, with GUT scale mass, which is more minimal, but behaves as CDM on galactic scales.
In section \ref{xiO1} we discuss the usual assumption of ``hidden charged dark matter" models, and show that if the dark sector has a temperature comparable to the SM sector, the hidden dark sector is required to possess interactions with the Standard Model via a mediator below the GUT scale. Finally,  we conclude in section \ref{conclusions}.

\section{Setup}\label{sec:setup}

We consider a dark matter particle that interacts only gravitationally with the SM, referred to as PIDM.
For concreteness, we focus on the case where the PIDM is a Dirac fermion $X$ with mass $m_X$.
In addition, we assume that $X$ is charged under a dark $U(1)$ gauge symmetry with gauge boson $\gamma_D$. 
For a Minkowski metric, the Lagrangian of the dark sector is given by
\beq
{\cal L}_{DM}= -\frac{1}{4} V_{\mu\nu}V^{\mu\nu} + \bar{X}i\slashed{D}X-m_X\bar{X}X\,,
\eeq
where  $V_{\mu}$ is the dark photon and $V_{\mu\nu}\equiv \partial_\mu V_\nu-\partial_\nu V_\mu$. Furthermore, $D_\mu = \partial_\mu - ig_D V_\mu$ is the covariant derivative, with $g_D$ the charge of the PIDM under the dark $U(1)$. We also define the dark fine-structure constant $4 \pi \alpha_D \equiv g_D^2$. For a general metric, we assume the dark sector to couple minimally to gravity.
In addition, we assume no direct couplings between the dark and SM sectors, leading to a maximal decoupling between them.
When expanding around a flat background $g_{\mu\nu}=\eta_{\mu\nu}+\sqrt{32\pi G} \, h_{\mu\nu}$ this leads to the total Lagrangian
\beq
{\cal L}={\cal L}_{SM}+{\cal L}_{EH}+{\cal L}_{DM}+\frac{\sqrt{8\pi}}{m_p}h^{\mu\nu}\left(T_{\mu\nu}^{SM}+T_{\mu\nu}^{DM}\right)\,,
\eeq
where the first two terms are the usual SM and the Einstein-Hilbert Lagrangians, $m_p=1/\sqrt{G}$ is the Planck scale,
$T_{\mu\nu}^{SM}$ is the energy-momentum tensor of the SM, and 
\beq
T^{\mu\nu}_{DM} = V^{\mu\rho}{V^\nu}_\rho-\frac14 \eta^{\mu\nu} V_{\rho\lambda}V^{\rho\lambda}+\left[\frac{i}{4}\bar X\left(\gamma^\mu D^\nu+\gamma^\nu D^\mu\right)X-\frac{1}{2}\eta^{\mu\nu}\bar X\left (i\slashed{D}-m_X\right)X
\,+\,{\rm h.c.}\right]\,,
\eeq
is the energy-momentum tensor of the dark sector.

Even if absent initially, direct couplings between the dark and visible sectors could be generated dynamically by quantum corrections. 
Possible operators need to be invariant under both the dark gauge symmetry $U(1)$, as well as the SM gauge group.
The former requirement prevents decays of a single $X$ to SM particles. Therefore, the $U(1)$ charge provides a mechanism to guarantee the stability
of the $X$ particle, making it absolutely stable. Note that this property constitutes a phenomenological difference compared to the uncharged PIDM, that could
decay with a long lifetime due to nonperturbative gravitational effects, with decay rate being exponentially suppressed \cite{Garny:2015sjg,Garny:2017kha}.

\begin{figure}
\centering
\includegraphics[width=8cm]{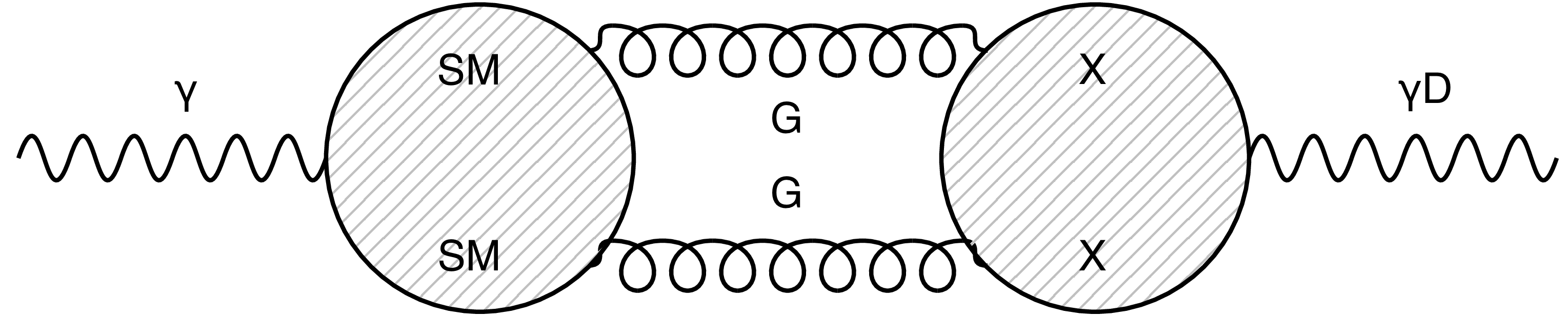}
\caption{\label{kinmix} The kinetic-mixing parameter $\epsilon$ could be generated through quantum corrections for which a photon is converted into a dark photon through loop diagrams involving SM particles, the PIDM, and gravitons. When starting from a purely gravitationally coupled theory, the loop corrections vanish for a Dirac fermion PIDM.
}
\end{figure}

A direct coupling at the renormalizable level arises from a possible kinetic mixing between the photon and the new gauge boson. Kinetic mixing is described by a dimension four operator in the Lagrangian of the form $\epsilon F_{\mu\nu} V^{\mu\nu}$, where $\epsilon$ is a dimensionless number. If $\epsilon$ is sizeable this  
operator could spoil our assumption that the two sectors are maximally decoupled. Even if we start from a Lagrangian with $\epsilon=0$, a non-zero kinetic mixing could be generated through loop corrections. In our scenario these loops have to involve gravitons, since this is the only particle that can communicate between the two sectors. 
The topology of possible Feynman diagrams that can contribute to the kinetic mixing is shown exemplarily in Fig.\,\ref{kinmix}. The photon ($\gamma$) couples to a charged SM particle, that is contained in the contribution to the loop amplitude indicated by the left shaded circle. The right circle contains at least one virtual PIDM line, with a coupling to the dark photon $\gamma_D$.
The parts of the amplitude contained within the two circles are connected by a number $N$ of graviton lines, where $N=2$ in Fig.\,\ref{kinmix}. 
The contributions with a single PIDM loop and $N=2$ vanish, analogously to the cancellation of the mixed $U(1)$/gravitational anomaly in the dark sector. The triangle diagram with one external dark photon and two external gravitons, which is the one giving rise to a gravitational anomaly, is proportional to $\mathrm{Tr} \, Q_D$, i.e. the sum of the $U(1)$ charges of the particles in the dark sector. This is trivially zero for a Dirac fermion PIDM, as the gauge anomalies cancel between the left-handed and right-handed spinors. Moreover, all corresponding diagrams with a single PIDM loop and $N$ external graviton lines are also proportional to $\mathrm{Tr} \, Q_D$, since the insertion of additional graviton lines does not change the structure of the electromagnetic coupling, and they also vanish. 

This argument can be generalized to contributions involving higher-loop corrections. 
In particular, in absence of non-gravitational interactions between the two sectors, the Lagrangian is invariant under charge conjugation
symmetry in the dark sector, for which $V_\mu\to-V_\mu$, $X\to X^c=i\gamma^2\gamma^0\bar X^T$, while all SM particles transform trivially.
The kinetic mixing term is odd under this symmetry, which means that it cannot be generated by loop effects, with $\epsilon$ transforming as a scalar.\footnote{Note that this conclusion would change
if the theory would encompass additional, heavy particles charged under both $U(1)$ symmetries. In this case the separate charge conjugation
symmetry is explicitly broken, and the theory would be invariant only under a common $C$ operation for which $A_\mu\to-A_\mu$, $V_\mu\to-V_\mu$. The same is true if
the theory contains additional heavy particles that are charged only under either one of the $U(1)$ symmetries, and in addition interact with a common massive
gauge boson corresponding to a broken $SU(N)$ theory. In this case a non-zero contribution could be generated from diagrams similar to the one in Fig.\,\ref{kinmix},
with gravitons replaced by $SU(N)$ gauge bosons.}

Apart from kinetic mixing, quantum corrections could induce higher-dimensional operators.
Since the graviton is the only mediator between the two sectors, all such couplings are generically Planck suppressed. 
One example is the dimension five operator $\frac{c_5}{\Lambda}\bar{X} X H^{\dagger} H$, where $H$ is the Higgs field,  $c_5$ a dimensionless coefficient, and $\Lambda\sim m_p$ the suppression scale. The PIDM particle and antiparticle have zero net ``dark" charge and could thus annihilate into SM neutral particles via this operator. 
The corresponding annihilation cross section is parametrically suppressed by $(c_5E/m_p)^2$ at low energies $E$.
Naively, one might expect $c_5\sim{\cal O}(1)$. However, when starting from a purely gravitationally coupled theory, $c_5$ from graviton loops is further suppressed compared
to the naive expectation.
The leading loop correction that could induce $c_5$ is a box diagram involving two gravitons, a Higgs, and a PIDM.
The graviton couplings contribute a factor $1/m_p^4$. Apart from that, the only scales entering in the loop integral are the
external momenta and the $X$ and Higgs masses. The momentum-dependence of the gravitational coupling contributes factors of the
loop momentum in the numerator, rendering the integral divergent. Using dimensional regularization, and assuming the external momenta
and Higgs mass to be much smaller than $m_X$, yields the dimensional estimate $c_5={\cal O}(m_X/m_p)^3$.
This implies that the contribution to the annihilation cross section is negligible compared to the tree-level contribution from $s$-channel
graviton exchange, as expected. We therefore take the latter into account in the abundance calculation~\cite{Garny:2015sjg, Garny:2017kha}, but neglect $c_5$.
Another example for a dimension five operator that would mediate $X\bar X$
annihilation is the operator $\frac{c_5'}{\Lambda}V_{\mu\nu}\bar\psi\sigma^{\mu\nu}\psi$, where $\psi$ is a SM fermion. Similar arguments as above prompt us to neglect also $c_5'$.

\section{Phenomenology of the charged PIDM}\label{pheno}

Self-interacting dark matter (SIDM) could help to explain small-scale structure observations that are in tension with numerical simulations
of collisionless CDM, such as the core-cusp, too-big-to-fail and diversity problems. Putting aside for a moment the possibility that these
discrepancies are entirely due to baryonic physics, we discuss to what extent the charged PIDM could provide self-interactions of the required order
of magnitude. In our scenario, DM particles scatter elastically with each other through $2 \rightarrow 2$ interactions mediated by a massless photon-like particle. In order to address the small-scale problems, the scattering probability must be marginally non-negligible within galactic environments, corresponding to a cross section 
per unit mass of the order of
\beq \label{sigma/m}
\sigma/m_X \sim 1 \, \si{cm^2/g} \approx 2 \times 10^{-24} \si{cm^2/GeV}.
\eeq
In the $U(1)$ model, the scattering cross section is enhanced at low relative velocities as $1/v^4$.
The order of magnitude of the cross section can be estimated by
\beq\label{selfscattering}
\sigma/m_X = \frac{8 \pi \alpha_D^2}{m_X^3 v^4} \sim  1 \, \si{cm^2/g} \, \left(\frac{\alpha_D}{2.5 \times 10^{-3}}\right)^2 \, \left(\frac{100 \, \si{GeV}}{m_X}\right)^3 \, \left(\frac{300 \, \si{km/s}}{v}\right)^4,
\eeq
where $v$ is the typical velocity of dark matter particles in galaxies. We see that we can have a large enough $\sigma/m_X$ for $m_X \sim 100 \, \si{GeV}$ and $\alpha_D \sim 10^{-3}$. While it has been generally believed that the unbroken $U(1)$ model is too constrained by the known properties of clusters, galaxies and dwarfs to be a good solution to the small-scale problems of structure formation, a more conservative attitude was recently argued for in  \cite{Agrawal:2016quu}. 

Note that the DM velocity $v$ is set by the gravitational potential for virialized objects, so it is essentially model-independent at late times and determined only by the type of object under consideration (galaxy, cluster, etc.). This means that if we want charged dark matter to solve the core-cusp problem and similar small-scale structure discrepancies, its mass is preferred to be in the ballpark of $10^2\, \si{GeV}$, depending on the coupling $\alpha_D$. 
On the other hand, as we will see below, the typical dark matter velocity $v$ at early times depends on its relevant interactions.

The calculation of the PIDM abundance can be split into two parts: first, around the time of reheating, the dark sector is populated
(dominantly by freeze-in production), see section \ref{100GEV}.
The subsequent evolution depends on whether the dark sector thermalizes, and may or may not alter the initial PIDM abundance, depending on
the size of $\alpha_D$, see section \ref{sec:evolution}. The results are presented and discussed in section \ref{sec:abundance},
including also the abundance of dark photons.

The PIDM scenario with a dark matter mass near the GUT scale is impracticable if we want to satisfy Eq.\,\eqref{sigma/m}. On the other hand, if one assumes that the small-scale problems are not solved by self-interacting dark matter, we are free to consider the possibility of a charged GUT scale PIDM.
We briefly discuss this case in section \ref{GUT}.

\subsection{Production of dark sector particles} \label{100GEV}

Due to their small non-renormalizable coupling to the SM, the dark sector particles ($X$ as well as dark photons) are produced at the highest available energy scales
in cosmic history. In general, one can discriminate several production mechanisms: ``freeze-in'' production shortly after reheating 
\cite{Chung:1998ua,Chung:1998rq,Hall:2009bx}, 
production during reheating \cite{Giudice:2000ex},
production during inflation, and so-called gravitational production \cite{Kuzmin:1999zk, Chung:2001cb}. 
The latter is effective for 
a narrow mass range around $m_X\sim H_i$, where $H_i$ is the Hubble scale at the end of inflation.
Production \emph{during} inflation is possible for light bosonic degrees of freedom with mass below $H_i$, but mostly ruled
out from isocurvature constraints \cite{Akrami:2018odb}.
For the fermion PIDM and the conformally coupled dark gauge boson
this is not relevant. Production during reheating depends on the reheating dynamics and can be
very model-dependent. For concreteness, following \cite{Giudice:2000ex}, we assume a perturbative reheating scenario described by a constant equation of state $w_\phi$ during
reheating and a decay rate $\Gamma_\phi=\gamma^2 H_i$ of the inflaton, parameterized by a dimensionless parameter $\gamma\leq 1$.
The reheating temperature is given by
\be
 T_{rh} = \kappa_2\gamma(m_pH_i)^{1/2}\,,
\ee
where $\kappa_2=(45/(4\pi^3g_{rh}))^{1/4}\simeq 0.25$ for $g_{rh}\simeq 10^2$.
Finally, production after reheating, during radiation domination, can be viewed as ``freeze-in'' via non-renormalizable interactions,
being dominated by the highest available temperature $T_{rh}$. For perturbative reheating with $w_\phi\simeq 0$, and PIDM mass
either much larger or much smaller than $H_i$, production during and after reheating dominates~\cite{Garny:2015sjg, Garny:2017kha}.
The PIDM and dark photon densities $n_{X}=n_{\bar X}$ and $n_{\gamma_D}$
can then be expressed in terms of the dimensionless quantities $X_{X}\equiv (n_{X}+n_{\bar X}) a^3/T_{rh}^3$ and 
$X_{\gamma_D}\equiv (n_{\gamma_D}) a^3/T_{rh}^3$ and are given by
\bea\label{eq:abundance}
  X_{X} &=& \frac{2}{T_{rh}^3}\int_{a_i}^{a_f} da\frac{a^2}{H(a)}\langle\sigma v\rangle_{X\bar X\to {\rm SM}\,{\rm SM}'}(n_{X}^{eq})^2\,,\nn\\
  X_{\gamma_D} &=& \frac{1}{T_{rh}^3}\int_{a_i}^{a_f} da\frac{a^2}{H(a)}\langle\sigma v\rangle_{\gamma_D\gamma_D\to {\rm SM}\,{\rm SM}'}(n_{\gamma_D}^{eq})^2\,,
\eea
where $n_{X(\gamma_D)}^{eq}=g_{X(\gamma_D)}\int\frac{d^3p}{(2\pi)^3}(e^{E_{X(\gamma_D)}/T}\pm 1)^{-1}$, $H(a)$ is the Hubble rate, $a_i\equiv 1$ the scale-factor at the end
of inflation, and $T$ the temperature of the SM thermal bath. Furthermore $g_X=g_{\gamma_D}=2$.
The integral saturates for times shortly after reheating, and in practice one can therefore set $a_f\to\infty$.
In the light PIDM regime $T\gg m_X$ during the relevant phase of freeze-in production. In this limit one has
$n_X^{eq}=\frac{2\zeta(3)}{\pi^2}T^3$, while $n_{\gamma_D}^{eq}=\frac34\times\frac{2\zeta(3)}{\pi^2}T^3$ holds for any temperature.

The thermally averaged cross sections correspond to $s$-channel graviton exchange and involve a sum over all possible SM particles in the final
state. Due to the gravitational interaction the cross sections depend only on the spin, but not on the other quantum numbers.
Note that in practice the freeze-in production is dominated by the inverse process, i.e. the gain term in the Boltzmann equations for
the dark sector particles. Accordingly, we neglected the loss term to arrive at the expression above.
The relevant cross sections for the fermion PIDM and the massless dark gauge boson are given by\footnote{For the case $T\ll m_X$ in which $X$
are produced non-relativistically, one needs to include Sommerfeld enhancement due to exchange of the dark gauge boson, see section\,\ref{GUT}.
Sommerfeld enhancement is not relevant within the light PIDM regime, for which $T\gg m_X$ during freeze-in production.} \cite{Garny:2017kha}
\be
  \langle\sigma v\rangle_{X\bar X(\gamma_D\gamma_D) \to {\rm SM}\,{\rm SM}'} = N_0\langle\sigma v\rangle_{0}+N_{1/2}\langle\sigma v\rangle_{1/2}
  +N_1\langle\sigma v\rangle_{1}\,,
\ee
where $N_0=4$, $N_{1/2}=45$, $N_1=12$ are the number of scalar, fermion and vector degrees of freedom in the SM.
For $X\bar X \to {\rm SM}\,{\rm SM}'$ the thermally averaged cross sections are given by 
\begin{eqnarray}\label{eq:sigXX}
\langle\sigma v\rangle_{0}&=&\frac{ \pi m_X T}{2 m_p^4}\left[\frac{4}{5}\frac{T}{m_X} +\frac{1}{5}\frac{m_X}{T}-\frac{1}{5}\frac{m_X}{T}\frac{K_1^2}{K_2^2}+\frac{2}{5}\frac{K_1}{K_2}\right]\to \frac{2 \pi  T^2}{5 m_p^4}\,,\nonumber\\
\langle\sigma v\rangle_{1/2}=\langle\sigma v\rangle_{1}&=&\frac{4 \pi m_X T}{m_p^4}\left[\frac{6}{5}\frac{T}{m_X} +\frac{2}{15}\frac{m_X}{T}-\frac{2}{15}\frac{m_X}{T}\frac{K_1^2}{K_2^2}+\frac{3}{5}\frac{K_1}{K_2}\right]\to \frac{24 \pi  T^2}{5m_p^4}\,,
\end{eqnarray}
where the modified Bessel functions are evaluated at $m_X/T$, with $T=T_{SM}$ being the temperature of the SM thermal bath.
The expressions right of the arrow denote the limit for $T\gg m_X$, relevant for the light PIDM regime.
For $\gamma_D\gamma_D \to {\rm SM}\,{\rm SM}'$, the thermally averaged cross sections are
\begin{eqnarray}
\langle\sigma v\rangle_{0}&=&\frac{3 \pi T^2}{5 m_p^4}\,,\nonumber\\
\langle\sigma v\rangle_{1/2}&=&\langle\sigma v\rangle_{1} \ =  \frac{208 \pi T^2}{5m_p^4}\,.
\end{eqnarray}
This gives (assuming $T\gg m_X$)
\be\label{eq:sigvrelativistic}
  \langle\sigma v\rangle_{X\bar X \to {\rm SM}\,{\rm SM}'} = \frac{1376 \pi T^2}{5m_p^4}\,,\quad
  \langle\sigma v\rangle_{\gamma_D\gamma_D \to {\rm SM}\,{\rm SM}'} = \frac{11868 \pi T^2}{5m_p^4}\,.
\ee
The ratio of dark photon and PIDM abundance produced via freeze-in is therefore given by
\be
  n_{\gamma_D}/(n_{X}+n_{\bar X}) \simeq \frac{621}{256}\,.
\ee

In order to produce an appreciable amount of $X$ particles with a mass around $100 \, \si{GeV}$ by freeze-in we need a 
practically instantaneous reheating with $\gamma\simeq 1$ and temperature around $T_{rh}\sim 10^{-4} m_p$ \cite{Garny:2017kha}. 
At these temperatures particles will be produced relativistically, so the thermally averaged cross sections from Eq.\,\eqref{eq:sigvrelativistic}
can be used. In addition, almost instantaneous reheating implies $H(a)\simeq H_i/a^2$.
In this limit the integrals in \eqref{eq:abundance} can be performed analytically, and we obtain
\be
 X_X=\frac{11008\kappa_2^2\zeta(3)^2}{15\pi^3}\left(\frac{T_{rh}}{m_p}\right)^3 \simeq 2.1\left(\frac{T_{rh}}{m_p}\right)^3, \quad
 X_{\gamma_D}=\frac{8901\kappa_2^2\zeta(3)^2}{5\pi^3}\left(\frac{T_{rh}}{m_p}\right)^3 \simeq 5.2\left(\frac{T_{rh}}{m_p}\right)^3\,. \quad
\ee
If interactions within the dark sector can be neglected, the number densities will scale with the usual $1/a^3$ factor
after reheating (see below for the discussion of the validity of this assumption). 
In addition, following \cite{Giudice:2000ex}, we include a dilution factor $\sim 1/8$ to account for residual entropy
production after the end of reheating. Writing for the total PIDM density $n_X+n_{\bar X}\equiv n_{i,X}a^{-3}$ and $n_{\gamma_D}\equiv n_{i,\gamma_D}a^{-3}$
then gives for the ``initial'' densities produced via freeze-in (recall our convention $a_i=1$ at the end of inflation)
\be\label{niX}
  n_{i,X}\simeq 0.27\frac{T_{rh}^6}{m_p^3},\quad n_{i,\gamma_D}\simeq 0.65\frac{T_{rh}^6}{m_p^3}\,.
\ee
Thus, putting everything together, we obtain $n_{i,X}, n_{i,\gamma_D} \propto T_{rh}^6$.

\subsection{Evolution of dark sector particles} \label{sec:evolution}

Freeze-in produces a non-thermal distribution of dark sector particles, with initial abundances $n_{i,X}$ and $n_{i,\gamma_D}$
computed above, and momentum distributions $f_X(p)$ and $f_{\gamma_D}(p)$ peaked
around the typical energy scale $\langle p_X\rangle\sim\langle p_{\gamma_D}\rangle \sim T_{SM}$ of order the SM temperature.
However, the initial number densities produced via freeze-in are much smaller, by a factor of order $(T_{rh}/m_p)^3$, 
than would be the case for an equilibrium distribution at this temperature. This means freeze-in produces an underpopulated
distribution.

The dark sector will never come in thermal equilibrium with the SM sector because the gravitational interactions are too weak, 
but it could in principle equilibrate within itself.
Here we discuss the relevant interactions, and whether they can be sufficiently strong.
An overview is provided in Tab.\,\ref{tab:overview}.

Of particular importance for the dark matter abundance is whether or not annihilations $X\bar X\to \gamma_D\gamma_D$ within the dark
sector become relevant. This is generally the case within the hidden charged dark matter scenario discussed in \cite{Agrawal:2016quu},
but, as we will see, not necessarily for the charged PIDM discussed here. In addition, the abundance of dark gauge bosons is
relevant for bounds on the total relativistic energy density parameterized by $\Delta N_{\rm eff}=N_{\rm eff}-3.046$ from the cosmic microwave 
background (CMB) anisotropies \cite{Aghanim:2018eyx}.

\begin{table}[h!]
\centering
{\scriptsize
\begin{tabular}{|ll|cc|cc|}
\hline
  &  &  \multicolumn{2}{c|}{kinetic equilibrium}   &  \multicolumn{2}{c|}{chemical equilibrium}   \\
  &  &  DM only & DM+$\gamma_D$   &  DM-$\gamma_D$ conversion &  total number \\
\hline
  &        & $XX\leftrightarrow XX$ & $X\gamma_D\leftrightarrow X\gamma_D$ & $X\bar X\leftrightarrow\gamma_D\gamma_D$ & $X\bar{X} \leftrightarrow X \bar{X} \gamma_D$ \\
  &        & $X\bar X\leftrightarrow X\bar X$ &  & \multirow{2}{70pt}{$X\bar X\leftrightarrow \underbrace{B_{X\bar X}}_{\to \gamma_D\gamma_D}\gamma_D$} & $X \gamma_D \leftrightarrow X \gamma_D \gamma_D$ \\
\multicolumn{2}{|c|}{Processes} & & & & $\gamma_D\gamma_D \leftrightarrow X \bar{X} \gamma_D$\\
  &        & & & &  $X \gamma_D \leftrightarrow X \bar{X} X$\\
  &        & & & &  $X \bar{X} \leftrightarrow \gamma_D \gamma_D \gamma_D$\\
\hline
\multirow{2}{20pt}{${\cal O}(\sigma v)$}
  & ultra-rel. & $\frac{\alpha_D^2}{E^2}$ & $\frac{\alpha_D^2}{E^2}$ & $\frac{\alpha_D^2}{E^2}$ & see Eq.\,\eqref{eq:split} \\[2ex]
  & non-rel.   & $\frac{\alpha_D^2}{m_X^2v^4}$ & $\frac{\alpha_D^2}{m_X^2}$ & $S_{ann/rec}\left(\frac{\alpha_D}{v}\right)\times\frac{\alpha_D^2}{m_X^2}$ & - \\
\hline
\end{tabular}
}

\caption{\small Overview of relevant processes for establishing kinetic equilibrium and chemical equilibrium within the dark sector, respectively, and order of magnitude of
the cross section in the ultra-relativistic regime ($E\gg m_X$) and in the non-relativistic regime ($E\ll m_X$). The first column
corresponds to kinetic equilibrium among dark matter particles, and the second to complete \emph{kinetic} equilibrium \emph{within the dark sector}. 
The third column captures the conversion
of dark matter into dark gauge bosons, relevant for freeze-out, and the last column number-changing interactions that can establish complete thermal equilibrium \emph{within the dark sector}. The corresponding rate $\Gamma=n_X\sigma v$ for collinear emission of gauge bosons, $X\bar X\to X\bar X\gamma_D$, is parametrically enhanced above the naive expectation $\sigma v\sim \alpha_X^3/E^2$, and drives the thermalization process within the relativistic regime, see text for details. For the first column, the cross section corresponds to the ``momentum transfer cross section'' $\sigma_{tr}=\int d\Omega\frac{d\sigma}{d\Omega}(1-\cos(\theta))$. Due to Planck suppression interactions with the Standard Model particles are irrelevant for both kinetic and chemical equilibrium.}
\label{tab:overview}
\end{table}

\subsubsection{Relativistic regime}

For the light PIDM scenario discussed above, $T_{SM}\gg m_X$ during freeze-in production, implying an
initially relativistic non-thermal distribution with $\langle p_X\rangle\gg m_X$.
Due to cosmic expansion, the typical momentum drops below $m_X$ at some point. It turns out to be useful to
separately discuss the regimes for which $\langle p_X\rangle\gg m_X$ and, at later times, $\langle p_X\rangle\ll m_X$.
Here we start with the first case.

To gain some intuition, let us first discuss what would happen if interactions within the dark sector
would establish complete thermal equilibrium while being in the relativistic regime.
The number densities will in general change when approaching thermal equilibrium, but the energy density remains covariantly conserved. 
Equating the initial energy density $\rho_{i,X}+\rho_{i,\gamma_D} \simeq (n_{i,X}+n_{i,\gamma_D}) \langle E_i\rangle 
\simeq (n_{i,X}+n_{i,\gamma_D}) T_{rh} \simeq T_{rh}^7/m_p^3$ to the equilibrium energy density $\rho_{eq}=6T_{i,D}^4\pi^2/30$, we can relate $T_{i,D}$ to $T_{rh}$.
Solving $\rho_{i,X}+\rho_{i,\gamma_D} =\rho_{eq}$ we find that, in complete thermal equilibrium, the temperature of the dark sector scales as 
\be
T_{i,D} \sim T_{rh}^{7/4}m_p^{-3/4}\,.
\ee
In the relativistic regime $T_D=T_{i,D}/a$ and therefore the ratio of the temperatures is given by
\be\label{xifulleq}
  \xi = \frac{T_D}{T_{SM}} = \left(\frac{T_{rh}}{m_p}\right)^{3/4}\,\left(\frac{g_*(T_{SM})}{g_{rh}}\right)^{1/3}\,,
\ee
so that if $T_{rh}\sim 10^{-4} m_p$, we find $T_{i,D} \sim 10^{-7} m_p$ and $\xi = T_D/T_{SM} \sim 10^{-3}$.
(This ratio gets modified in the non-relativistic regime, see Eq.\,\eqref{eq:xiloweq}.)

The PIDM and dark gauge boson population produced shortly after reheating via freeze-in
can therefore be considered as an \emph{underoccupied} non-equilibrium initial distribution with typical particle energy given by the ``hard'' scale
\be\label{eq:Ehini}
 E_h\big|_{\rm ini} \simeq T_{rh}\,.
\ee
In absence of interactions the energies redshift, giving
\be
  E_h \simeq T_{rh}a^{-1}\,,
\ee
where we used $a_i=1$ at the end of inflation. We will see that interactions in the dark sector modify the time-dependence,
and therefore treat $E_h$ as a generic time-dependent quantity for the moment, with initial value \eqref{eq:Ehini}.
In the following we refer to the corresponding distribution functions $f_X^h=f_{\bar X}^h$ and
$f_{\gamma_D}^h$ as ``hard'' particles, with number densities 
\be
n_X^h=n_{\bar X}^h=g_X\int\frac{d^3p}{(2\pi)^3}f_X^h\,,
\ee
and $n_{\gamma_D}^h$ defined analogously. Initially, $n_X^h|_{\rm ini}=n_{i,X}$ and analogously for $\bar X$
and $\gamma_D$. In absence of interactions affecting the distribution and total number density of hard particles
one has $f_X^h(a,p)=f_X^h(1,ap)$ and $n_X^h=n_{i,X}a^{-3}$.

\begin{centering}
	\begin{figure*}[th]
		\centering
		\includegraphics[width=.4\textwidth]{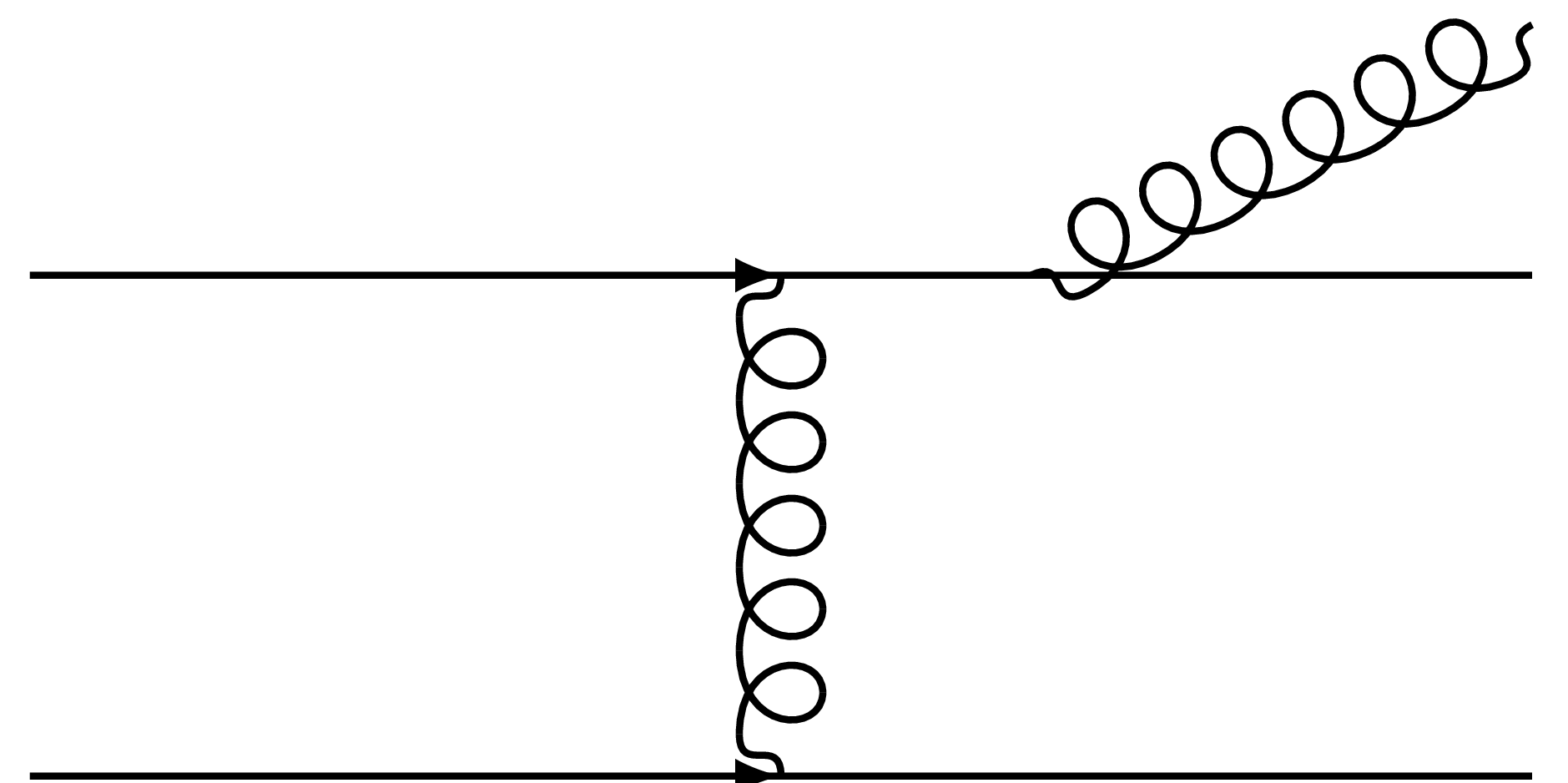}
		\caption{The bremsstrahlung process leading to the potential thermalization of the dark sector.}
		\label{reheatdiagrams}
	\end{figure*}
\end{centering} 

The dark gauge interactions lead to two types of interactions within the dark sector: elastic two-to-two scatterings
and inelastic processes such as bremsstrahlung, see Fig.~\ref{reheatdiagrams}. 
In the following we present parametric estimates of various processes
relevant for thermalization in the dark sector in the relativistic regime, where the PIDM mass can be neglected compared
to $\langle p_X\rangle$. We ignore various logarithmic corrections, and keep only the leading power dependence in
all estimates in order to establish the scaling of the thermalization rate, depending on the parameters
 $\alpha_D$ and $T_{rh}/m_p$, assuming $\gamma\simeq 1$. For small $\alpha_D$ thermalization, if it takes place at all, occurs on time-scales that
are long compared to the Hubble rate. Therefore, the expansion has to be taken into account.

\paragraph{Elastic $2\to 2$ scattering:}

In the relativistic regime, elastic scatterings with an ${\cal O}(1)$
momentum transfer can establish kinetic equilibrium. For example, the rate for $X\bar X\to X\bar X$ is given by
\be
  \Gamma_{el,tr} = n_X^h \sigma_{tr} v \simeq \frac{\alpha_D^2n_X^h}{E_h^2}\,,
\ee
where $\sigma_{tr}=\int d\Omega\frac{d\sigma}{d\Omega}(1-\cos\theta)$ is the so-called momentum transfer cross section
relevant for elastic scatterings with ${\cal O}(1)$ momentum transfer. In the relativistic regime, the related
scattering processes $X\gamma_D\to X\gamma_D$ (dark Compton scattering) and $X\bar X\to\gamma_D\gamma_D$ (pair annihilation)
occur at comparable rates. The rate $\Gamma_{el,tr}$ can become larger than the Hubble rate while $E_h\gg m_X$ if
\be\label{eq:alphacritel}
  \alpha_D \gg \alpha_D^{\rm crit,el} \equiv 
  \left(\frac{m_X}{\kappa_2^2 m_p}\right)^{1/2}\left(\frac{T_{rh}^3}{n_{i,X}}\right)^{3/2}
  \simeq 10^{-2}\,\left(\frac{m_X}{100\,{\rm GeV}}\right)^{1/2} \left(\frac{10^{-4}m_p}{T_{rh}}\right)^{3/2}\,.
\ee
This means that, if this condition is satisfied, elastic scatterings can become relevant for the thermalization process.
As we will see, for realistic values $T_{rh}\ll 10^{-3}m_p$ this is \emph{not} the case.
Instead, inelastic processes are more efficient for thermalization, leading to a weaker condition on $\alpha_D$ for thermalization to occur.

\paragraph{Inelastic scattering processes:}

Naively, one might think that inelastic processes are suppressed, because the cross section for kinematically allowed $2\to 3$ processes involves an
additional vertex compared to $2\to 2$ scatterings, potentially leading to a relative suppression by a factor of $\alpha_D$.
However, it is well known that for both Abelian and non-Abelian gauge interactions this is not the case.
The basic physical reason can be roughly understood in the following way (we provide a more detailed discussion below, and ignore logarithmic corrections, as stated above): 
even though $1\to 2$ processes such as $X\to X\gamma_D$ are kinematically forbidden,
they become allowed if either the parent or daughter particle is slightly off-shell. This can occur due to a $2\to 2$ scattering before or after
the $1\to 2$ splitting. In contrast to the elastic case discussed above, also $2\to 2$ scatterings with small momentum transfer contribute to the
effective, combined $2\to 3$ process. If the associated $2\to 2$ scattering proceeds via Coulomb scattering with a dark gauge boson in the $t$-channel,
this leads to a parametric enhancement of the relevant rate $\Gamma_{el}\simeq \alpha_D^2n_X^h/\mu^2$ as compared to $\Gamma_{el,tr}$ by
a factor of order $E_h^2/\mu^2$, where $\mu$ is the IR cutoff of the $t$-channel momentum exchange (see below). In fact, this enhancement factor may over-compensate
the suppression factor $\alpha_D$ associated to the additional $1\to 2$ splitting, leading in total to a larger rate as compared to elastic scatterings. 

\begin{centering}
	\begin{figure*}[th]
		\centering
		\includegraphics[width=.5\textwidth]{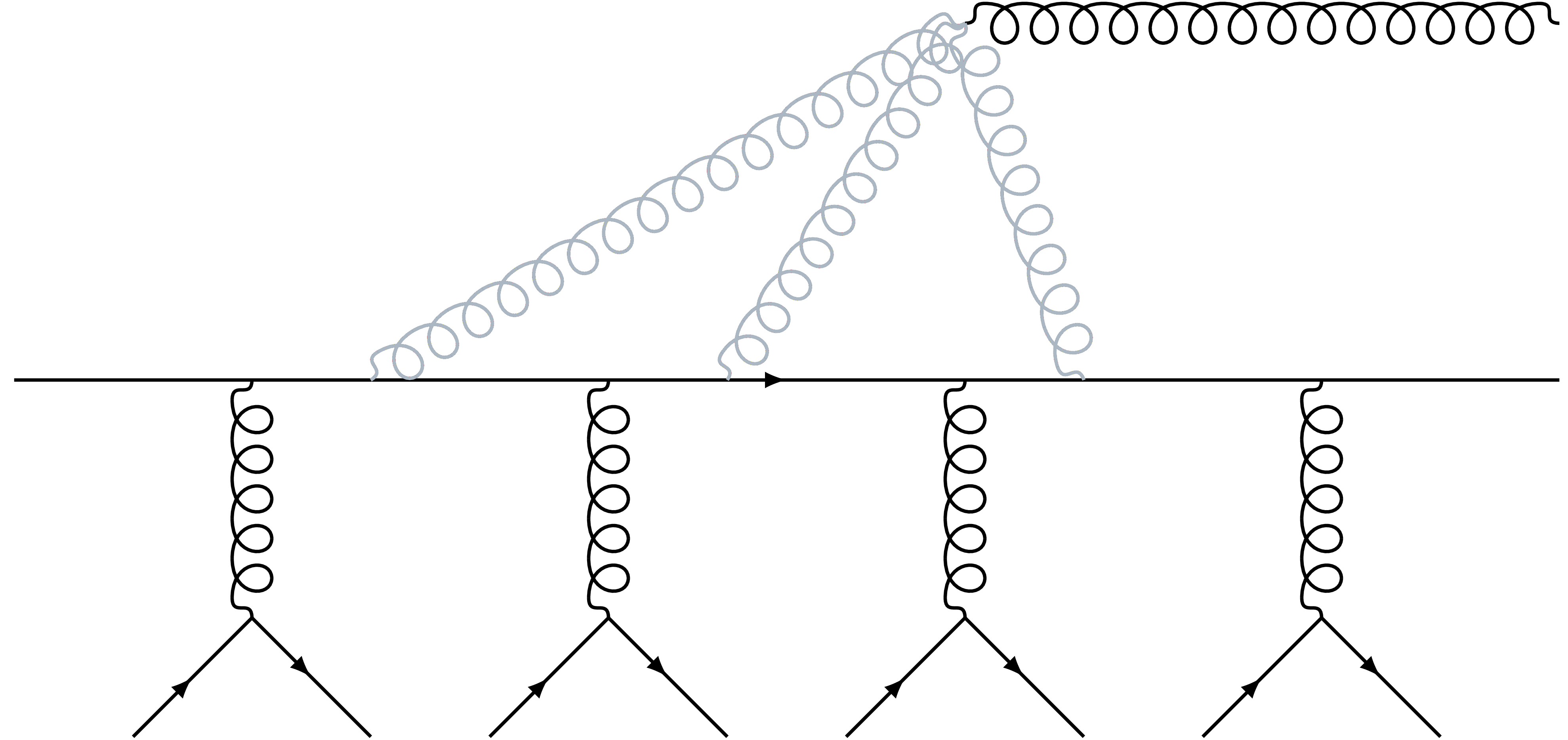}
		\caption{The LPM process, leading to a suppression of the dark sector thermalization rate for high enough density.  The grey propagators are to be understood as being summed over in separate diagrams, which serve to partially interfere to suppress the full amplitude.}
		\label{reheatdiagrams2}
	\end{figure*}
\end{centering}

For very large $\Gamma_{el}$ this simple picture gets modified, because
several $2\to 2$ scatterings can occur during the time-scale of the $1\to 2$ splitting, known as Landau-Pomeranchuk-Migdal (LPM) effect. This reduces the
total inelastic scattering rate compared to the case where all $2\to 2$ events can be treated as independent from each other. Technically, it requires
to resum contributions to the amplitude of successive $2\to 2$ scatterings, with the gauge boson being radiated off any of the intermediate propagators,
see Fig.\,\ref{reheatdiagrams2}.
The LPM effect can be seen as a destructive interference, leading to a suppression. Nevertheless, as we will see, inelastic processes can dominate over
elastic ones.

For the case of a non-Abelian gauge interaction, both elastic and inelastic processes have been described by an effective kinetic theory setup \cite{Arnold:2002zm} 
that has been used to describe the initial stages of the thermalization process in the context of relativistic heavy-ion collisions \cite{Kurkela:2011ti,Kurkela:2014tea},
in the weakly coupled limit.  It has also been applied to thermalization within the SM after inflation for reheating with a very small inflaton decay rate \cite{Mukaida:2015ria}. Here we apply this setup to the relativistic, underoccupied population of
$X$, $\bar X$ and $\gamma_D$ produced via freeze-in to describe the subsequent evolution after reheating.

The splitting rate $X(p)\bar X\to X(p')\bar X\gamma_D(k,\theta)$ for a dark gauge boson $\gamma_D$ with momentum $k\lesssim p$ emitted under a (small)
angle $\theta$ relative to the momentum $\vec p$ of the incoming $X$ can be estimated by \cite{Mukaida:2015ria}
\be\label{eq:split}
  \Gamma_{\rm split}(k,\theta) \simeq \Gamma_{el}\, k\frac{dI_{\rm rad}}{dk} \simeq \Gamma_{el}\,\alpha_D\,{\rm min}\left[1,\frac{\Gamma_{\rm form}(k,\theta)}{\Gamma_{el}}\right]\,,
\ee
where $\Gamma_{el}\equiv\frac{\alpha_D^2n_X}{\mu^2}$ is the relevant Coulomb scattering rate of a hard primary $X$ off a charge (here $\bar X$)
with transverse momentum transfer $q_\perp^2\gtrsim \mu^2$, where $\mu^2$ is an IR cutoff (usually of the order of the Debye mass, see below). Here $q$
is the momentum of the $t$-channel exchange, and $\vec q_\perp$ is the component perpendicular to $\vec p$, and we assume $p^0\sim |\vec p|\sim E_h$
 as well as $|q_\mu|\ll E_h$.
Destructive interference between individual scatterings reduces the rate if $\Gamma_{\rm form}<\Gamma_{el}$.
This LPM suppression can be qualitatively understood in terms of the following picture:
the emission of the gauge boson requires a certain time-scale, the so-called formation time $\Delta t_{\rm form}$ related to the
virtuality of order $\kappa^2\equiv (k+p')^2\simeq (k+p)^2\sim k\cdot p \sim kE_h\theta^2$ of the intermediate $X$ particle. Boosted to the frame of the
hard primary, the corresponding time-scale is $\Delta t_{\rm form} \sim \frac{E_h}{\kappa}\,\kappa^{-1}\sim \frac{1}{k\theta^2}=\frac{k}{k_\perp^2}$,
where $k_\perp\equiv k\theta$ is the transverse momentum of the gauge boson.
If the ``formation'' time-scale for the emitted gauge boson is longer than the time $\Gamma_{el}^{-1}$ between two scatterings, only a single
photon will be emitted during that period.

For an Abelian gauge interaction, the typical range of angles is related to the angle of the outgoing $X(p')$, 
$\theta\sim \theta_X' \sim q_\perp/E_h$, which gives $k_\perp=k\theta\sim kq_\perp/E_h$ and $\Delta t_{\rm form}\sim \frac{E_h^2}{k q_\perp^2}$ \cite{Peigne:2008wu}.
The formation time-scale therefore depends on the typical values of $q_\perp^2$. Inside the medium, a large number of elastic scatterings
occurs during the formation process, and one needs to take the typical distribution of $q_\perp^2$ into account. A simple physical picture
can be obtained by viewing the subsequent scatterings as random contributions to the transverse momentum, leading to a diffusion process
for which the mean squared-value $\langle q_\perp^2\rangle$ increases linearly with time. In order to properly take the expansion of the
universe into account we switch to conformal time $d\tau=dt/a$ and express all momenta in terms of comoving momenta as $k_{\rm com}=ka$.
The diffusion process can be described by \cite{Mukaida:2015ria}
\be
  \frac{d}{d\tau}\langle q_{\perp,{\rm com}}^2\rangle = \hat q_{\rm com}\,,
\ee
where the diffusion constant $\hat q_{\rm com}=a^3\hat q$ is given by
\be
  \hat q \sim \int d^2q_{\perp} \frac{\partial \Gamma_{el}}{\partial q_\perp^2}q_\perp^2 \sim \alpha_D^2 g_X\int\frac{d^3p}{(2\pi)^3} f_X(p)(1-f_X(p))\sim \alpha^2 n_X\,,
\ee
where we neglected Pauli blocking in the last step and used the estimate
\be
  \frac{\partial \Gamma_{el}}{\partial q_\perp^2} \sim \frac{\alpha_D^2}{q_\perp^2(q_\perp^2+\mu^2)}g_X\int\frac{d^3p}{(2\pi)^3} f_X(p)(1-f_X(p))\,.
\ee
In conformal time the formation time-scale $\Delta\tau_{\rm form}\sim \frac{E_{h,{\rm com}}^2}{k_{\rm com} q_{\perp,{\rm com}}^2}$ is thus due to
the diffusion process on average given by $\Delta\tau_{\rm form}\sim \sqrt{\frac{E_{h,{\rm com}}^2}{k_{\rm com} \hat q_{com}}}$. The corresponding
formation rate, with respect to physical time, is
\be
  \Gamma_{\rm form} \sim \frac{1}{a\Delta\tau_{\rm form}} \sim \sqrt{\frac{k \hat q}{E_{h}^2}}\,,
\ee
where we have expressed all quantities in terms of physical momenta again.\footnote{In the non-Abelian case the emitted gauge boson can
itself couple to the virtual gauge boson mediating the scattering, leading to a typical angle $\theta\sim q_\perp/k$, i.e. $k_\perp\sim q_\perp$. In this case 
one obtains $\Gamma_{\rm form} \sim \sqrt{\frac{\hat q}{k}}$, agreeing with Eq. (3.10) in \cite{Mukaida:2015ria}. This leads to a different modification of the
shape of the bremsstrahlung spectrum for low $k$ due to the LPM effect ($\propto 1/\sqrt{k}$ instead of $\sqrt{k}$ for the Abelian case), but the same parametric dependence for $k\lesssim E_h$ relevant for the total number density $n_{\gamma_D}^s$, see below.}.

The IR cutoff $\mu$ is usually related to the Debye screening scale $m_D$, which can for a non-equilibrium distribution be estimated as
\be\label{eq:mD}
  m_D^2 \sim \alpha_D g_X\int\frac{d^3p}{(2\pi)^3}\frac{f_X(p)}{p} \sim \alpha_D n_X/T_*\,,
\ee
where, in the last step, we introduced the effective ``temperature'' \cite{Kurkela:2014tea}
\be\label{eq:Tstar}
  T_* \equiv \frac12 \int\frac{d^3p}{(2\pi)^3}f_X(p)(1-f_X(p)) \Bigg/ \int\frac{d^3p}{(2\pi)^3}\frac{f_X(p)}{p}\,.
\ee
Neglecting Pauli blocking the numerator is related to the total number density $n_X$.
Initially we expect $T_*\sim E_h \sim T_{rh}a^{-1}$ to be of the order of the ``hard'' energy scale.

In order to avoid unrealistically large IR contributions from momentum transfer $q_\perp^2<H^2$ we assume $\mu={\rm max}(m_D,H)$.
Altogether, the rate for gauge boson production is parametrically given by
\be\label{eq:Gammasplit}
  \Gamma_{\rm split} \sim \left\{\begin{array}{ll}
    {\rm min}\left[\alpha_D^2T_*,\alpha_D^2\sqrt{\frac{kn_X}{E_h^2}}\right] & m_D>H \\
    {\rm min}\left[\frac{\alpha_D^3n_X}{H^2},\alpha_D^2\sqrt{\frac{kn_X}{E_h^2}}\right] & m_D<H
  \end{array}\right. \qquad (m_X\ll\mu)\,.
\ee
This estimate is valid in the ultra-relativistic limit, when the PIDM mass $m_X$ is negligible.
We comment on the effect of $m_X$ below.
Initially, shortly after reheating, the case $m_X,m_D\ll H$ is realized. In addition, for $T_{rh}/m_p\gtrsim 10^{-5}$ and $k\lesssim E_h$
the expression involving the `min' function in Eq.\,(\ref{eq:split}) evaluates to unity for times shortly after reheating, i.e. there is (initially) no LPM suppression. In this regime
the ratio of the inelastic to the elastic rate is given by
\be
  \Gamma_{\rm split}/\Gamma_{el,tr}\Big|_{ini} \sim \frac{\alpha_DE_h^2}{H^2} \sim \kappa_2^4\alpha_D\frac{m_p^2}{T_{rh}^2}
\ee
Therefore, inelastic processes dominate immediately after reheating if the dark fine-structure constant satisfies 
the rather weak condition $\alpha_D\gtrsim 10^{-6}(T_{rh}/(10^{-4}m_p))^2$. As we will see below, inelastic processes dominate also at later
times and therefore drive the thermalization process.

\begin{centering}
	\begin{figure*}[th]
		\centering
		\includegraphics[width=.8\textwidth]{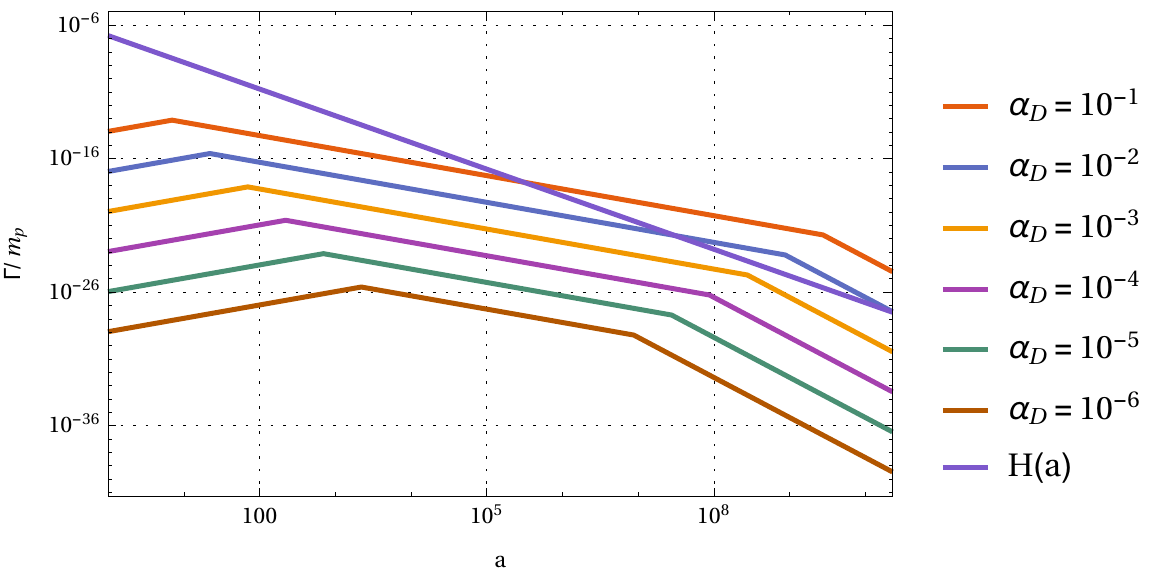}
		\caption{The evolution of $\Gamma_\text{split}$ with scale-factor $a$ within the relativistic regime
                for $T_{rh}=10^{-4}m_p$, $m_X=100$\,GeV and various values of the dark fine-structure constant.
                Thermalization within the dark sector requires $\Gamma_\text{split}>H$,
                and the Hubble expansion rate $H$ is also shown for comparison. We use the normalization $a=1$ at the end of inflation.
                For each value of $\alpha_D$, $\Gamma_\text{split}\sim \alpha_D^3n_X/\mu^2$ is for early times given by Eq.\,\eqref{eq:Gammasplit} with
                $m_X,m_D\ll H=\mu$. The first kink occurs when LPM suppression sets in, and after that
                $\Gamma_\text{split}\sim \alpha_D^2\sqrt{\frac{kn_X}{E_h^2}}$ (we assume $k\lesssim E_h$ in the figure).
                Subsequently, the Debye scale $m_D$ becomes larger than $H$, which however does not affect the splitting rate in the LPM
                regime, see Eq.\,\eqref{eq:Gammasplit}. Next, $m_D$ drops below the PIDM mass $m_X$. As discussed in the text, due to an
                interplay of the IR cutoff for Coulomb scattering and the formation time this
                also does not affect the parametric scaling of the splitting rate in the LPM regime, see Eq.\,\eqref{eq:Gammasplit2}.
                The second kink occurs once $\alpha_DE_h(n_X/k)^{1/2}$ drops below $m_X^2$, see Eq.\,\eqref{eq:Gammasplit3}. 
                At this point LPM suppression stops and $\Gamma_\text{split}\sim\alpha_D^3\frac{n_X}{m_X^2}$ decreases faster than the Hubble rate $H$.
                }
		\label{gammasplit}
	\end{figure*}
\end{centering}

So far we neglected the PIDM mass $m_X$ in the discussion and assumed all particle species are ultra-relativistic.
Let us now comment on modifications due to the finite mass. As before, we are interested only in the power law dependence
on the various energy scales and neglect logarithmic modifications. By assumption, we consider the relativistic
regime $m_X\ll E_h$ in this section. Eq.\,\eqref{eq:Gammasplit} is correct as long as the (stronger)
condition $m_X\ll \mu$ holds, i.e. when the mass is negligible compared to the IR cutoff. For $m_X\gg \mu$
but not too large (see below) the bremsstrahlung spectrum $k\frac{dI_{\rm rad}}{dk}$ becomes suppressed by a multiplicative factor 
of order $\mu^2/m_X^2$ \cite{Peigne:2008wu}. In addition, in the LPM regime, the typical angle is now $\theta\sim\theta_m= m_X/E_h$ which increases the
virtuality of the intermediate state and therefore reduces the formation time. It turns out that this compensates for the suppression
of the spectrum \cite{Peigne:2008wu}, such that the splitting rate can be estimated as
\be\label{eq:Gammasplit2}
  \Gamma_{\rm split} \sim 
    {\rm min}\left[\alpha_D^3\frac{n_X}{m_X^2},\alpha_D^2\sqrt{\frac{kn_X}{E_h^2}}\right] \qquad \left(\mu^2\ll m_X^2 \ll \alpha_DE_h(n_X/k)^{1/2}\right)\,.
\ee
For even larger mass, when $\Delta t_{\rm form, heavy}\equiv (k\theta_m^2)^{-1}= E_h^2/(k m_X^2)<\Delta t_{\rm form}$, which occurs for 
$m_X^2\gtrsim \alpha_DE_h(n_X/k)^{1/2}$, the LPM effect does not occur any more, and
\be\label{eq:Gammasplit3}
  \Gamma_{\rm split} \sim 
    \alpha_D^3\frac{n_X}{m_X^2} \qquad \left(\alpha_DE_h(n_X/k)^{1/2} \ll m_X^2 \ll E_h^2\right)\,.
\ee
Note that, in all cases, the splitting rate is parametrically larger than the naive estimate of order $\alpha_D^3n_X/E_h^2$, which
would be correct for a $2\to 3$ process in which the momentum exchange between the charged particles is of order $E_h$.  The evolution of $\Gamma_{\rm split}$ is depicted in Fig.~\ref{gammasplit}.

In order to discuss the time evolution, one needs to take into account the secondary population of dark sector particles
produced by the collinear splitting rate $\Gamma_{\rm split}$ discussed above. Since their momenta are distributed below
the hard scale, $k\lesssim E_h$, we refer to them as ``soft'' particles, with distribution function $f^s$ and number density $n^s$.
The production of soft gauge bosons can be described by the Boltzmann equation (we suppress the time argument in all functions of $k$) \cite{Mukaida:2015ria}
\be
  (\partial_t-Hk\partial_k)f_{\gamma_D}^s(k) \sim \Gamma_{\rm split}(k)n_X^hk^{-3}\,.
\ee
Up to logarithmic corrections this implies for the corresponding number density $n_{\gamma_D}^s$
\be\label{eq:ns}
  \frac{1}{a^3}\frac{d}{dt}(a^3n_{\gamma_D}^s) \sim \Gamma_{\rm split}(E_h)n_X^h\,.
\ee
To obtain this equation we used an upper cutoff for the integration over $k$ given by $k_{\rm max}\sim E_h$.
Note that, when ignoring logarithmic corrections, the lower cutoff $k_{\rm min}$ in $k$ does not appear in this 
parametric estimate. In practice, it should be given either by the IR cutoff $\mu={\rm max}(m_D,H)$, 
or by some scale $E_s$ below which rescattering of soft gauge bosons establishes a softer
thermal distribution $f^s\propto k^{-1}$ \cite{Kurkela:2011ti, Mukaida:2015ria}. 
Since the ``soft'' population is by definition produced via inelastic interactions \emph{after} reheating, 
their initial abundance vanishes, $n_{\gamma_D}^s|_{\rm ini}=0$.

In addition, also a population of soft $X$ and $\bar X$ is produced by related processes.
In particular, for the pair creation process $\gamma_DX\to X\bar X X$, the photon may first split into an $X\bar X$
pair, with either $X$ or $\bar X$ being  slightly off-shell, and scattering off another $X$ particle, similar to the process
discussed above. Furthermore, for the soft particles, dark Compton and annihilation processes occur at rates enhanced by factors
of $E_h/k$ and $(E_h/k)^2$, respectively, as compared to corresponding scatterings among the hard particles. We therefore assume
for simplicity that, provided dark gauge interactions become relevant at all, they are efficient enough to produce a
distribution of soft $X$ and $\bar X$ with comparable number density, $n_X^s= n_{\bar X}^s\sim n_{\gamma_D}^s$. This
assumption should be scrutinized in a more detailed treatment, going beyond the scope of the present work.

Note that the collinear splitting rate \eqref{eq:Gammasplit} involves the total number density $n_X=n_X^s+n_X^h$.
Therefore, an efficient production of a soft bath of particles further enhances their production.
In addition, their contribution to the IR sensitive integral in the numerator in \eqref{eq:Tstar} will tend to decrease $T_*$.
Parametrically, using an IR cutoff $k_{\rm min}\sim E_s$ as discussed before, \eqref{eq:Tstar} yields
\be
  (T_*)^{-1} \sim \frac{n_X^h}{n_X}E_h^{-1} + \frac{n_X^s}{n_X}E_s^{-1}\,.
\ee
where $E_s$ is the ``soft'' energy scale below which the distribution thermalizes.
If full thermalization in the dark sector is reached we expect
\be
E_s \sim T_D = \xi T_{SM}
\ee
to be given by the appropriate dark sector temperature \eqref{xifulleq}.
At early times $T_*\to E_h\sim T_{SM}$ for $n_X^s\to 0$, as discussed previously, and at late times $T_*\to E_s \sim T_D$
and $n_X^s\gg n_X^h$.

The typical energy of the hard population is diminished by radiative processes,
with energy loss rate given by (for comoving momenta and conformal time) $dE_{h,{\rm com}}/d\tau\sim - k_{\rm com}\Gamma_{\rm split,com}(k)$ \cite{Peigne:2008wu},
where $\Gamma_{\rm split,com}=a\Gamma_{\rm split}$. In terms of physical momenta and time $t$,
the energy loss rate is
\be
\frac{1}{a}\frac{d}{dt}(aE_{h}) \sim - k\Gamma_{\rm split}(k)\,.
\ee
In absence of interactions $E_h=T_{rh}/a\propto a^{-1}$, as discussed before.
When interactions become relevant, processes with $k\lesssim k_{\rm max}=E_h$ dominate, such that
\be
  E_h \sim T_{rh}a^{-1}\,\exp\left(-\int_{t_i}^t dt' \Gamma_{\rm split}(E_h(t'))\right)\,.
\ee
Following \cite{Kurkela:2011ti}, we assume the hard population is thermalized at time $t_h$ when $E_h$ has dropped to the equilibrium temperature $T_D=\xi T_{SM}$
of the dark sector. This gives the following condition for the time $t_h$,
\be\label{eq:th}
  \xi \overset{!}{=}  \exp\left(-\int_{t_i}^{t_h} dt' \Gamma_{\rm split}(E_h(t'))\right)
   = \exp\left(-\int_{0}^{\eta_h} d\eta' \frac{\Gamma_{\rm split}(E_h(\eta'))}{H(\eta')}\right)\,,
\ee
where we introduced $\eta\equiv\ln(a)$. Note that the total number density of \emph{hard} particles scales as $n_X^h=n_{iX}a^{-3}$ for $t<t_h$
since their number does not change by any of the relevant interactions, at leading power in $\alpha_D$. On the other
hand, their energy decreases as described above. 

Due to the exponential suppression, once $\Gamma_{\rm split}/H>1$ the hard particles quickly lose their energy and radiate particles with $k\lesssim E_h$.
Since an ${\cal O}(1)$ fraction of the ``daughter'' population itself has momenta of order of (but smaller as) $E_h$, they also rapidly lose
energy by a number of subsequent scatterings within a Hubble time $\Delta t=H^{-1}$. The further secondary radiation leads to an increase in
the soft population. The secondary processes can be estimated parametrically by replacing $n_X^h\to n_X=n_X^h+n_X^s$ on the right-hand side of \eqref{eq:ns},
i.e. considering the  ${\cal O}(1)$ fraction of ``soft'' secondaries with $k\sim {\cal O}(1)\,E_h$ as a source for further bremsstrahlung. The quick exponential decrease of $E_h$ then
leads to a corresponding increase in $n^s\equiv n_{\gamma_D}^s\sim n_X^s$. This increase should stop when all particles have lost their energy and rescatterings in the dark sector
lead to thermal equilibrium at some time $t_s$. For simplicity, we estimate that thermalization has occurred once  the soft population $n^s$ reaches
the equilibrium number density within the dark sector $n_{eq}\sim T_D^3\sim \xi^3T_{\rm SM}^3$ with $\xi$
given in \eqref{xifulleq}. We assume that by the time this occurs, interactions within the dark sector are strong enough
to maintain equilibrium such that $n^s=n_{eq}$ for $t>t_s$. For the parameter range we are interested in, it turns out that $n^s\ll n_{eq}$
for $t\ll t_h$. When $t\to t_h$ the density $n_s$ increases with an exponential factor related to the one in \eqref{eq:th} (with positive sign in the exponential). 
Consequently, $t_s$ is of the same order as $t_h$, up to logarithmic corrections that we systematically neglect.

Following the previous discussion, the thermalization time can be estimated analytically, again up to logarithmic corrections, by the condition that 
the ratio $\Gamma_{\rm split}/H$ reaches unity\footnote{This condition is analogous to the thermalization time-scale being $\Gamma_{\rm split}^{-1}$
for the case of an initially underoccupied, isotropic, non-expanding, weakly coupled 
$SU(N)$ plasma discussed in \cite{Kurkela:2011ti}. However, in that case, the contribution of the secondary, soft population further enhances
$\Gamma_{\rm split}$ (i.e. ``catalyzes'' the inelastic scattering rate). This is not the case here. This can be attributed to several differences,
in particular the three-dimensional expansion, which dilutes the number density.}. As long as this ratio is less than one, one has $n_X\simeq n_X^h=n_{i,X}a^{-3}$ and $E_h\sim T_{rh}a^{-1}$. 
Here we describe the relative size of the relevant quantities determining $\Gamma_{\rm split}$.
In the following parametric estimates we use that the momentum $k$ of the radiated gauge boson carries a typical momentum
fraction of order one, i.e. $k\sim {\cal O}(1)\,E_h$.
Immediately after reheating, $\mu=H$, and LPM suppression is irrelevant. After some expansion,
the formation time increases and LPM suppression sets in (corresponding to the first kink when following the evolution of
the splitting rate for a given value of $\alpha_D$ in Fig.\,Fig.~\ref{gammasplit}), such that 
\be
  \Gamma_{\rm split}(k\lesssim E_h)\sim \alpha_D^2\sqrt{\frac{n_X}{E_h}}\,.
\ee 
In this period $\Gamma_{\rm split}\propto a^{-1}$ drops slower than the Hubble rate $H\propto a^{-2}$, and therefore the ratio
$\Gamma_{\rm split}/H$ increases with time (straight segment between the two kinks in Fig.~\ref{gammasplit}).
At some point, $m_D$ becomes larger than $H$, such that $\mu=m_D$, and then $m_X$ becomes larger than $\mu$. However, due to the LPM effect, this does not
affect the splitting rate, as long as the condition $m_X^2 \ll \alpha_D(n_XE_h)^{1/2}$ holds. Once this condition is violated (second kink 
in Fig.~\ref{gammasplit}), the splitting rate drops as $\Gamma_{\rm split}\propto a^{-3}$, i.e. \emph{faster} than the Hubble rate $H\propto a^{-2}$. Therefore, a necessary condition for thermalization is that $\Gamma_{\rm split}/H$ becomes larger than unity \emph{before} $\alpha_D(n_XE_h)^{1/2}$ drops below $m_X^2$.
This can be converted in a condition on the dark gauge coupling, giving
\be
  \alpha_D\gg \alpha_D^{\rm crit,inel,(a)}\equiv \left(\frac{m_X}{\kappa_2^2 m_p}\right)^{2/5}\left(\frac{T_{rh}^3}{n_{i,X}}\right)^{3/10}
  \simeq 2\cdot 10^{-3}\,\left(\frac{m_X}{100\,{\rm GeV}}\right)^{2/5} \left(\frac{10^{-4}m_p}{T_{rh}}\right)^{9/10}\,.\label{acrit}
\ee

Since we consider the relativistic regime in this section, and all estimates are based on the assumption that the typical momentum 
satisfies $\langle p_X\rangle\gg m_X$, an additional condition is that $\Gamma_{\rm split}/H$ reaches unity when the temperature $T_D=\xi T_{SM}$ 
corresponding to full equilibrium in the dark sector is still much larger than $m_X$. This gives the condition
\be
  \alpha_D\gg \alpha_D^{\rm crit,inel,(b)}\equiv \left(\frac{m_X}{\kappa_2^2\xi m_p}\right)^{1/2}\left(\frac{T_{rh}^3}{n_{i,X}}\right)^{1/4}
  \simeq 5\cdot 10^{-4}\,\left(\frac{m_X}{100\,{\rm GeV}}\right)^{1/2} \left(\frac{10^{-4}m_p}{T_{rh}}\right)^{9/8}\,.
\ee
For thermalization to occur, both conditions must be satisfied, i.e. the combined condition on the dark fine-structure constant reads
\be\label{eq:alphacrit}
\alpha_D\gg \alpha_D^{\rm crit}\equiv{\rm max}(\alpha_D^{\rm crit,inel,(a)},\alpha_D^{\rm crit,inel,(b)})\,.
\ee
As noted before, inelastic processes are more efficient than elastic scattering, and therefore give a \emph{lower} value for the required
coupling strength than the estimate based on the $2\to 2$ rate (see Eq.\,\eqref{eq:alphacritel}). This means that for any value of the
coupling larger than $\alpha_D^{\rm crit}$, at the time when $\Gamma_{\rm split}/H$ reaches unity, $\Gamma_{el,tr}/H\ll 1$ is still
strongly suppressed. In addition, elastic processes, even if relevant, would not increase the total number of particles in the dark
sector, and therefore would not lead to full (chemical and kinetic) equilibration. Since thermalization is driven by inelastic processes,
the dark sector fully equilibrates, with temperature $T_D=\xi T_{SM}$ and vanishing chemical potentials, if the condition \eqref{eq:alphacrit} is satisfied.
For the benchmark scenario $m_X\sim 100$\,GeV and $T_{rh}/m_p\sim 10^{-4}$, thermalization within the dark sector thus occurs
before $X$ becomes non-relativistic for $\alpha_D\gg 10^{-3}$. On the other hand, interactions within the dark sector play no role for $\alpha_D\ll 10^{-4}$,
and the number densities and typical energy are given by the redshifted freeze-in values.
Within the intermediate range, thermalization is still incomplete once $X$ starts to become non-relativistic. We do not attempt to model this transition
region in this work.

\subsubsection{Non-relativistic regime}\label{sec:nonrelregime}

Once the typical momentum of PIDM particles drops below their mass, they become non-relativistic with typical velocities $v_X\ll 1$.
Furthermore, their \emph{equilibrium} abundance becomes Boltzmann suppressed.
The dark gauge boson is massless and remains relativistic.
For $\alpha_D\gg \alpha_D^{\rm crit}$, the dark sector is equilibrated at temperature $T_D=\xi T_{SM}$, while for $\alpha_D\ll \alpha_D^{\rm crit}$
the distribution is the redshifted initial distribution produced via freeze-in with typical momentum $p_X\sim T_{rh}/a$.
Therefore, the transition to the non-relativistic regime occurs for (we assume $a_i=1$ at reheating)
\be
  a_{nr} \simeq \left\{\begin{array}{ll}
  \frac{\xi T_{rh}}{m_X} \simeq 10^{10}\left(\frac{100\,{\rm GeV}}{m_X}\right)\left(\frac{T_{rh}}{10^{-4}m_p}\right)^{7/4} & \alpha_D\gg \alpha_D^{\rm crit}\,, \\
  \frac{T_{rh}}{m_X} \simeq 10^{13}\left(\frac{100\,{\rm GeV}}{m_X}\right)\left(\frac{T_{rh}}{10^{-4}m_p}\right) & \alpha_D\ll \alpha_D^{\rm crit}\,.
  \end{array}\right.
\ee

The question we are mostly interested in is whether annihilations $X\bar X\to \gamma_D\gamma_D$ reduce the abundance of PIDM, leading
to a freeze-out in the dark sector.
Since the annihilation cross section depends on the typical velocity $v_X$, we also need to consider its evolution, which is
determined by Compton scattering $X\gamma_D\to X\gamma_D$ within the dark sector.
Finally, due to the strong velocity dependence, self-interactions ($X\bar X\to X\bar X$, $XX\to XX$) may affect the momentum
distribution. 

Inelastic processes are less relevant in the nonrelativistic regime \cite{Peigne:2008wu}.
The reason is two-fold: one the one hand, if $\alpha_D\gg \alpha_D^{\rm crit}$, the system is already thermalized.
The subsequent evolution is then sensitive to processes dropping \emph{out} of equilibrium, such as the freeze-out of $X\bar X$ annihilation.
For that process, $2\to 2$ annihilation gives the dominant contribution.
On the other hand, for $\alpha_D\ll \alpha_D^{\rm crit}$, the inelastic rate is already smaller than the Hubble rate at the beginning
of the non-relativistic regime. In the following we assume that, in this case, it remains below $H$ also within the nonrelativistic regime, such that full thermalization does not occur.  Below, we discuss the various relevant reactions case by case.

\paragraph{Scattering $X\gamma_D\to X\gamma_D$:}

This type of scattering corresponds to Compton scattering in the dark sector,
and tends to equilibrate the dark gauge boson and PIDM temperatures.
If dark Compton scattering is \emph{not} in equilibrium within the non-relativistic regime, 
the typical momenta $p_X$ and $p_{\gamma_D}$ both scale as $1/a$ due to cosmic expansion.
This leads to the velocity scaling $v_X\propto 1/a$. If Compton scattering is efficient enough to
establish kinetic equilibrium within the dark sector, $T_X=T_{\gamma_D}\propto 1/a$, such that
the typical velocity scales as $v_X=\sqrt{3T_X/m_X} \propto 1/\sqrt{a}$.

In the non-relativistic limit the cross section is given by the Thomson cross section 
\beq
  \sigma_T = \frac{8\pi\alpha_D^2}{3m_X^2}\,.
\eeq
The scattering leads to a transfer of energy between the PIDM and gauge boson population.
Taking cosmic expansion into account as well, the rate of change of the PIDM temperature is given by \cite{Peebles:1994xt}
\be
  \frac{1}{a^2}\frac{d}{dt}\left(a^2T_X\right) = \frac{8\sigma_T\rho_{\gamma_D}}{3m_X}\,(T_X-T_{\gamma_D})\,.
\ee
The temperature equilibrates if $\Gamma_{\rm kin}\equiv 8\sigma_T\rho_{\gamma_D}/(3m_X)\gg H$.
Since $\Gamma_{\rm kin}\propto a^{-4}$, the ratio $\Gamma_{\rm kin}/H$ decreases with time.
If $\alpha_D\gg \alpha_D^{\rm crit}$, one has $\rho_{\gamma_D}=\frac{\pi^2}{15}T_D^4$, and $\Gamma_{\rm kin}/H>1$
for $a<a_{\rm kin}$, with
\be\label{eq:akin}
  a_{\rm kin} \simeq 3\cdot 10^{13}\,\left(\frac{\alpha_D}{10^{-2}}\right)\left(\frac{100\,{\rm GeV}}{m_X}\right)^{3/2}\left(\frac{T_{rh}}{10^{-4}m_p}\right)^{5/2} \qquad (\alpha_D\gg \alpha_D^{\rm crit})\,.
\ee
If $\alpha_D\ll \alpha_D^{\rm crit}$ one can check that $\Gamma_{\rm kin}/H$ is below unity for all times $a>a_{nr}$ and $T_{rh}\lesssim 10^{-3}m_p$,
such that Compton scatterings play no role within the range of possible reheating temperatures. This implies for the typical PIDM velocity,
\be\label{eq:vX}
  v_X \simeq \left\{\begin{array}{ll}
  (a/a_{nr})^{-1/2} & \alpha_D\gg \alpha_D^{\rm crit}, a_{nr}<a<a_{\rm kin} \\
  (a/\sqrt{a_{\rm kin}a_{nr}})^{-1} & \alpha_D\gg \alpha_D^{\rm crit}, a>a_{\rm kin} \\
  (a/a_{nr})^{-1} & \alpha_D\ll \alpha_D^{\rm crit}, a_{nr}<a 
  \end{array}\right.
\ee
while $v_X\sim 1$ for $a<a_{nr}$. The typical velocity will be important for $X\bar X$ annihilation, which we turn to next.

Note that for a non-Abelian gauge symmetry dark Compton scatterings $X\gamma_D \to X\gamma_D$ are enhanced at low momentum transfer \cite{Buen-Abad:2015ova}, similar to dark Coulomb scattering. This would enhance the cross section and extend the range over which kinetic equilibrium between the PIDM and the dark gauge bosons holds.
We do not discuss this possibility any further here.

\paragraph{Annihilation $X\bar X\to \gamma_D\gamma_D$:}

If the annihilation of PIDM particles is efficient in the non-relativistic regime, it leads to
a freeze-out within the dark sector which has an important impact on the final abundance.
The tree-level annihilation cross section is of order $\sigma_{\rm ann} v \sim \pi\alpha_D^2/E_X^2$.
The relevant quantity for freeze-out is the average over the distribution function $f_X(p)=f_{\bar X}(p)$ of PIDM (anti-)particles,
\beq
  \langle\sigma_{\rm ann} v\rangle 
  = \frac{ \int\frac{d^3p}{(2\pi)^3}\frac{d^3p'}{(2\pi)^3} \, f_X(p) f_{\bar X}(p') \, \sigma_{\rm ann} v }{ \int\frac{d^3p}{(2\pi)^3}\frac{d^3p'}{(2\pi)^3} \, f_X(p) f_{\bar X}(p') }\,.
\eeq
As discussed before, for small $\alpha_D\ll \alpha_D^{\rm crit}$ the distribution function is not necessarily given by a thermal distribution.
Nevertheless, due to the production via freeze-in, we expect the dominant contribution to come from
particles with $p_X\sim T_{SM}$ in this case. For simplicity, we use a thermal distribution function for
estimating the averaged cross section in that case. We checked that this provides a valid estimate
of the order of magnitude by comparing to a narrowly peaked distribution with the same average energy, see App.\,\ref{appendix}.
For $\alpha_D\gg \alpha_D^{\rm crit}$, the distribution is thermal with temperature $T_D=\xi T_{SM}$.

For relative velocities $v\ll 1$, exchange of dark gauge bosons leads to Sommerfeld enhancement \cite{Baer:1998pg, Hisano:2002fk, Hisano:2003ec}.
In addition, $X\bar X$ pairs may form bound states that then annihilate \cite{vonHarling:2014kha}.
The Sommerfeld enhanced annihilation rate is given by 
\be
\Gamma_{\rm ann} \equiv \Gamma(X \bar{X} \rightarrow \gamma_D \gamma_D)= n_X \langle\sigma_{\rm ann} v \times {S}_{\rm ann}\rangle \,,
\ee 
where $\sigma_{\rm ann} v$ is the tree-level cross section for $X \bar{X} \rightarrow \gamma_D \gamma_D$, $v$ the relative (M\o{}ller) velocity of the dark matter particles and
\beq\label{sommerfeld}
S_{\rm ann}(\alpha_D/v) = \frac{2 \pi \alpha_D/v}{1-e^{-2 \pi \alpha_D/v}}\,,
\eeq
is the Sommerfeld enhancement factor. For $v\ll 1$ the tree-level cross section $\sigma_{\rm ann} v\sim \pi\alpha_D^2/m_X^2$ approaches a constant $s$-wave
limit and can be pulled out of the average. As discussed above, the average  is computed assuming a Maxwell-Boltzmann velocity 
distribution, $f_X(p)\propto e^{-p^2/(2m_XT_X)}$. The averaged enhancement factor $\bar{S}_{\rm ann}\equiv \langle S_{\rm ann}\rangle$ 
is given by
\beq\label{avsomm}
\bar{S}_{\rm ann}= \frac{x^{3/2}}{2 \sqrt{\pi}} \int_0^{\infty} S_{\rm ann}(\alpha_D/v) v^2 e^{-\frac{xv^2}{4}} dv\,,
\eeq
where $x\equiv m_X/T_X$.

\begin{figure}[th]
\begin{centering}
\includegraphics[width=14cm]{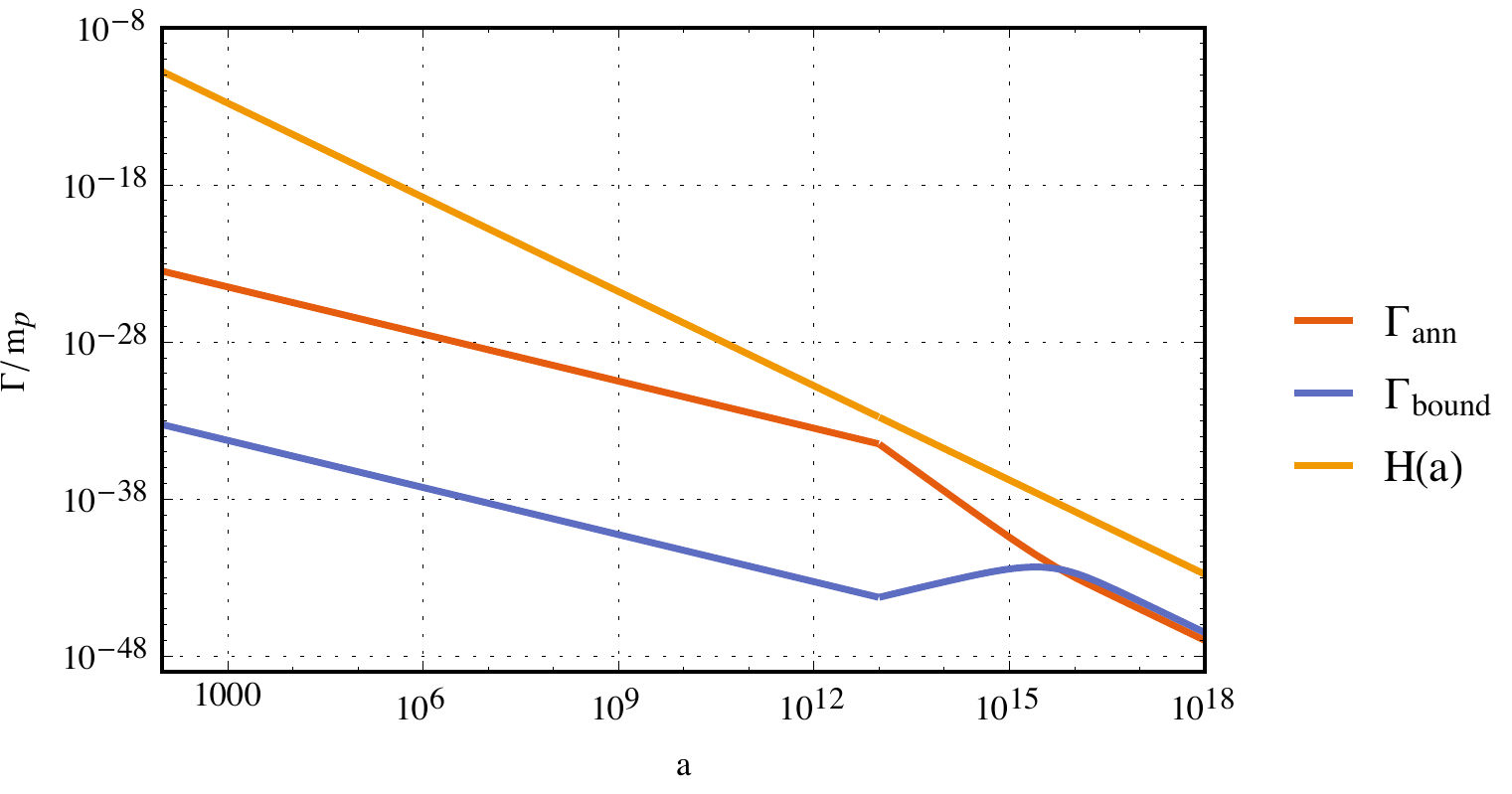}
\caption{\small Comparison of the Hubble expansion rate $H(a)$ (yellow), the annihilation rate $\Gamma_{\rm ann}$ (red) and bound state creation rate $\Gamma_{\rm rec}$
(blue) with the respective enhancement factors as a function of the scale factor for $\alpha_D = 0.001$, $m_X = 100 \, \si{GeV}$, $T_{rh}= 10^{-4}m_p$. 
Annihilation within the dark sector is negligible as long as the coupling constant is weak enough such that the dark sector never reaches thermal equilibrium, $\alpha_D \lesssim \alpha_D^{\rm crit}$. Here $a=1$ corresponds to reheating, while $a=10^{13}$ is the value for which $X$ becomes non-relativistic.}
\label{boundstate}
\end{centering}
\end{figure}

Let us first discuss the case $\alpha_D\ll \alpha_D^{\rm crit}$. In this case $\langle p_X\rangle \simeq T_{rh}/a$, which implies $T_X\simeq T_{rh}^2/(a^2m_X)$,
i.e. $x\simeq (a/a_{nr})^2$. At the beginning of the non-relativistic regime $x\sim 1$, i.e. $S_{\rm ann}  \sim 1$. Using $n_X=n_{i,X}a^{-3}$ one can check
that $\Gamma_{\rm ann}\sim n_X \pi \alpha_D^2/m_X^2$ is below $H$ for $a\gtrsim a_{nr}$ and possible values of $T_{rh}$.
Let us now check whether Sommerfeld enhancement can boost the annihilation rate to become larger than $H$ in the non-relativistic regime $a\gg a_{nr}$,
even when the coupling satisfies the above inequality.
In the non-relativistic limit, $S_{\rm ann}  \sim \pi \alpha_D/v$ and therefore $\Gamma_{\rm ann}\sim n_X \pi^2 \alpha_D^3/(m_X^2 v_X)$. The annihilation rate is enhanced by the factor $\pi \alpha_D/v_X$. 
This enhancement is still not enough to overtake the Hubble expansion rate, as is shown in Fig. \ref{boundstate}, so the relation $\Gamma_{\rm ann} \ll H$ still holds. In particular, using \eqref{eq:vX}, implies that after $X$ becomes non-relativistic the annihilation rate and the Hubble rate have the same dependence on the scale factor (in the logarithmic plot of Fig. \ref{boundstate} the two curves are parallel to each other after that point). Therefore, if annihilations are inefficient when $X$ becomes non-relativistic, they are also inefficient later on. We can therefore conclude that for values of the coupling constant $\alpha_D \ll \alpha_D^{\rm crit}$, annihilation is negligible at all times.
 
The rate of bound state formation can be estimated by $\Gamma_{\rm rec} \equiv \Gamma(X \bar{X} \rightarrow B \gamma_D)=n_X \langle\sigma_{\rm rec} v\rangle$. 
To a good approximation, the bound state creation cross section is just the annihilation cross section enhanced by a ``recombination factor'', i.e. 
$\langle\sigma_{\rm rec} v\rangle = \bar{S}_{\rm rec} \langle\sigma_{\rm ann} v\rangle$. Setting $\zeta \equiv \alpha_D/v$, the enhancement factor is \cite{vonHarling:2014kha}
\beq
\label{eq:boundstate}
S_{\rm rec} = S_{\rm ann}(\zeta) \frac{2^9}{3} \frac{\zeta^4}{(1+\zeta^2)^2} e^{-4 \zeta \cot^{-1} \zeta}.
\eeq
The thermally averaged recombination factor is defined as in Eq.\,\eqref{avsomm} by integrating over the Maxwell-Boltzmann velocity distribution. For large $v$, where $\zeta$ is close to zero, $S_{\rm rec} \ll S_{\rm ann}$ and the bound state effect becomes negligible. On the other hand, in the small kinetic energy limit, the two enhancement factors are comparable. The rate $\Gamma_{\rm rec}$ is also shown in Fig.\,\ref{boundstate}.

The conclusion is that for $\alpha_D \lesssim \alpha_D^{\rm crit}$ the annihilation rate within the dark sector is below the Hubble expansion rate both in the relativistic and non-relativistic regimes.
 For $\alpha_D \gg \alpha_D^{\rm crit}$, on the other hand, annihilation can occur. We will discuss how the resulting freeze-out in the dark
sector affects the final $X$ abundance in Sec.\,\ref{sec:abundance}.

\paragraph{Scattering $X\bar X\to X\bar X$, $XX\to XX$:}

Even if the complete dark sector cannot establish equilibrium for $\alpha_D \ll \alpha_D^{\rm crit}$, 
PIDM self-scatterings $X \bar{X} \rightarrow X\bar{X}$ and $XX\to XX$ via dark photon exchange (``dark Coulomb scattering'')
can bring the PIDM alone in kinetic equilibrium in the non-relativistic regime, leading to a Maxwell-Boltzmann distribution $f_X$ with a non-zero (negative) chemical potential. This is due to the fact that the self-scattering cross section is enhanced by $1/v_X^4$ (Eq.\,\eqref{selfscattering}), which becomes large at later epochs when $v_X$ is very small, see \eqref{eq:vX}. 

The scattering rate is given by $\Gamma\simeq n_X\sigma v_X$. Using\,\eqref{selfscattering} for the momentum-transfer cross section, and, 
for $\alpha_D \ll \alpha_D^{\rm crit}$, $n_X=n_{i,X}a^{-3}$ and \eqref{eq:vX} for $v_X$, one finds
\be
  \frac{\Gamma}{H} \sim \left(\frac{\alpha_D}{10^{-4}}\right)^2\left(\frac{100\,{\rm GeV}}{m_X}\right)\left(\frac{T_{rh}}{10^{-4}m_p}\right)^3
  \left(\frac{a}{50 a_{nr}}\right)^2 \qquad (\alpha_D \ll \alpha_D^{\rm crit}, a>a_{nr})\,.
\ee
Thus, even for a very small coupling $\alpha_D$, self-scatterings become efficient quickly after the beginning of the non-relativistic regime, for $a/a_{nr}\gtrsim 50$ for the benchmark values used for the normalization above.

As the velocity continues to drop, the self-interaction cross section can increase to enormous values. At the epoch of matter-radiation equality, the velocity is of the order of $10^{-10}$ for the benchmark scenario, corresponding to an enhancement factor for $\sigma v$ of $10^{30}$. At some point, the dark matter particles start to virialize in halos and $v_X$ increases again.  It is reasonable to ask whether there is some cutoff that shuts off the Coulomb enhancement $1/v^4$ below some critical velocity. The plasma contains charged particles that will screen electrostatic effects at distances larger than the Debye length $\lambda_D \simeq \sqrt{T_*/(n_X \alpha_D)}$, where $n_X$ is the number density of non-relativistic PIDM particles and $T_*\simeq m_X$  for $\alpha_D \ll \alpha_D^{\rm crit}$ and $a>a_{nr}$ (see Eq.\,\ref{eq:mD}). This effectively corresponds to a ``mass'' for the dark gauge boson, 
\beq\label{debyem}
m_{D}\simeq \sqrt{\frac{n_X \alpha_D}{m_X}} \simeq 0.5\,\frac{T_{rh}^3}{m_p^{3/2}} \sqrt{\frac{\alpha_D}{m_X}} a^{-3/2},
\eeq
which scales as $a^{-3/2}$, unlike the velocity. In the regime where $m_{D} \gg m_X v_X$ and $m_D\gg m_X\alpha_D$, scattering is a contact interaction and the momentum-transfer cross section becomes velocity independent, capped at the value $\sim \alpha_D^2 m_X^2/m_{D}^4$. For $m_X\alpha_D \gg m_{D} \gg m_X v_X$, and to the extent the Debye screening can be characterized by a mass term, non-perturbative effects similar to Sommerfeld enhancement can play a role \cite{Tulin:2012wi,Tulin:2013teo}.
For $\alpha_D \ll \alpha_D^{\rm crit}$, using $n_X=n_{i,X}a^{-3}$ and \eqref{eq:vX} one finds
\be
  \frac{m_D}{m_Xv_X} \simeq 10^{-8}\left(\frac{\alpha_D}{10^{-4}}\right)^{1/2}\left(\frac{T_{rh}}{10^{-4}m_p}\right)^{3/2}\,\times\,\left(\frac{a}{a_{nr}}\right)^{-1/2}\,.
\ee
Therefore, Debye screening is never important for $a>a_{nr}$ and cannot act as a cutoff in this scenario.

We conclude that even for a weak coupling $\alpha_D \ll \alpha_D^{\rm crit}$
the DM distribution function changes during its evolution, evolving from a non-thermal distribution $f_X^{NE}(a,p)$ peaked around $p\sim T_{SM}\approx T_{rh}/a$ to an equilibrium distribution with a negative chemical potential $f_X^E(a,p) \propto \exp (-\frac{E_X-\mu_X}{T_{X}})$, with temperature $T_X\simeq T_{rh}^2/(m_Xa^2)$. The transition occurs
somehwat after the time when $X$ becomes non relativistic, depending on the size of $\alpha_D$.
Note that the change in $f_X$ could in principle also alter the previous estimate of the averaged annihilation rate. 
However, as mentioned previously, the precise shape turns out to have only a minor impact as long as the average
momentum is parametrically the same (see App.\,\ref{appendix}). This condition turns out to be satisfied in the present case: the Maxwell-Boltzmann
distribution with temperature $T_X\propto 1/a^2$ in the non-relativistic regime corresponds to typical momenta $p\sim T_{rh}/a$, of the same order
of magnitude as for the redshifted initial distribution produced via freeze-in.

\subsection{Final abundance of dark matter and dark photons}\label{sec:abundance}

The conclusion of the discussion in the previous section is that the
phenomenology of the charged PIDM depends on whether the dark fine-structure
constant is smaller or larger than the critical value $\alpha_D^{\rm crit}$
defined in Eq.\,\eqref{eq:alphacrit}. In the following we discuss both cases in turn.

\subsubsection{Weak coupling regime $\alpha_D\ll \alpha_D^{\rm crit}$:}

In this regime the final dark matter abundance is set exclusively by freeze-in, while
interactions within the dark sector play a minor role. One notable exception is dark matter
self-interaction, that affects the shape of the dark matter distribution function,
and can play a role for structure formation.

The dark matter abundance \eqref{niX} obtained from freeze-in translates into a density parameter
\beq
 \Omega_X h^2 \simeq 0.12 \left(\frac{m_X}{390\,{\rm GeV}}\right) \left(\frac{T_{rh}}{6\cdot 10^{-4}m_p}\right)^3\,.
\eeq
The bound $r<0.064$ (95\% C.L.) on the tensor-to-scalar ratio \cite{Akrami:2018odb} translates into a conservative upper bound $T_{rh}/m_p\lesssim 6\cdot 10^{-4}$.
Requiring that all of the observed dark matter abundance $\Omega_c h^2=0.120\pm0.001$ \cite{Aghanim:2018eyx} is composed of PIDM therefore
requires a mass of at least $m_X\gtrsim 400$\,GeV. This value can be slightly lowered due to uncertainties in the production during reheating,
and the residual entropy production after reheating. Assuming instantaneous transition to radiation domination without residual entropy
production increases $\Omega_X h^2$ by a factor $8$ and correspondingly decreases the lower mass bound to $50$\,GeV.

In addition, the abundance of dark gauge bosons is also fixed by the freeze-in computation \eqref{niX}.
This can be translated into an energy density assuming a typical energy $E_{\gamma_D}\sim T_{rh}$ at reheating,
giving today
\beq
  \rho_{\gamma_D,0} \simeq T_{rh} n_{i,\gamma_D}(T_0/T_{rh})^4(g_*(T_0)/g_{rh})^{4/3}\simeq 0.01\rho_{\gamma,0}\left(\frac{T_{rh}}{m_p}\right)^3\,.
\eeq
Alternatively, one can express this extra radiation density in terms of a contribution to the ``effective number of neutrino species'',
\beq\label{eq:Neff}
  \Delta N_{\rm eff} \simeq 0.052\left(\frac{T_{rh}}{m_p}\right)^3\,.
\eeq
For allowed values of $T_{rh}$ this contribution is safely within the allowed range $ N_{\rm eff} =2.99\pm 0.17$ from
CMB \cite{Aghanim:2018eyx}.
The low value of $\Delta N_{\rm eff}$ discriminates this scenario from ``hidden charged dark matter'' setups in which the dark sector is initially in
thermal equilibrium with the SM, such that $\Delta N_{\rm eff}\geq 0.054$. This amount of extra radiation will be probed
by future CMB and large-scale structure observations \cite{Baumann:2015rya, Baumann:2017gkg}. Therefore, if dark matter is composed of a particle charged under an unbroken dark gauge force,
but no extra radiation is found in the future, this would point towards the PIDM scenario (see also Sec.\,\ref{xiO1}).

\begin{figure}[th]
\begin{centering}
\includegraphics[width=0.475\textwidth]{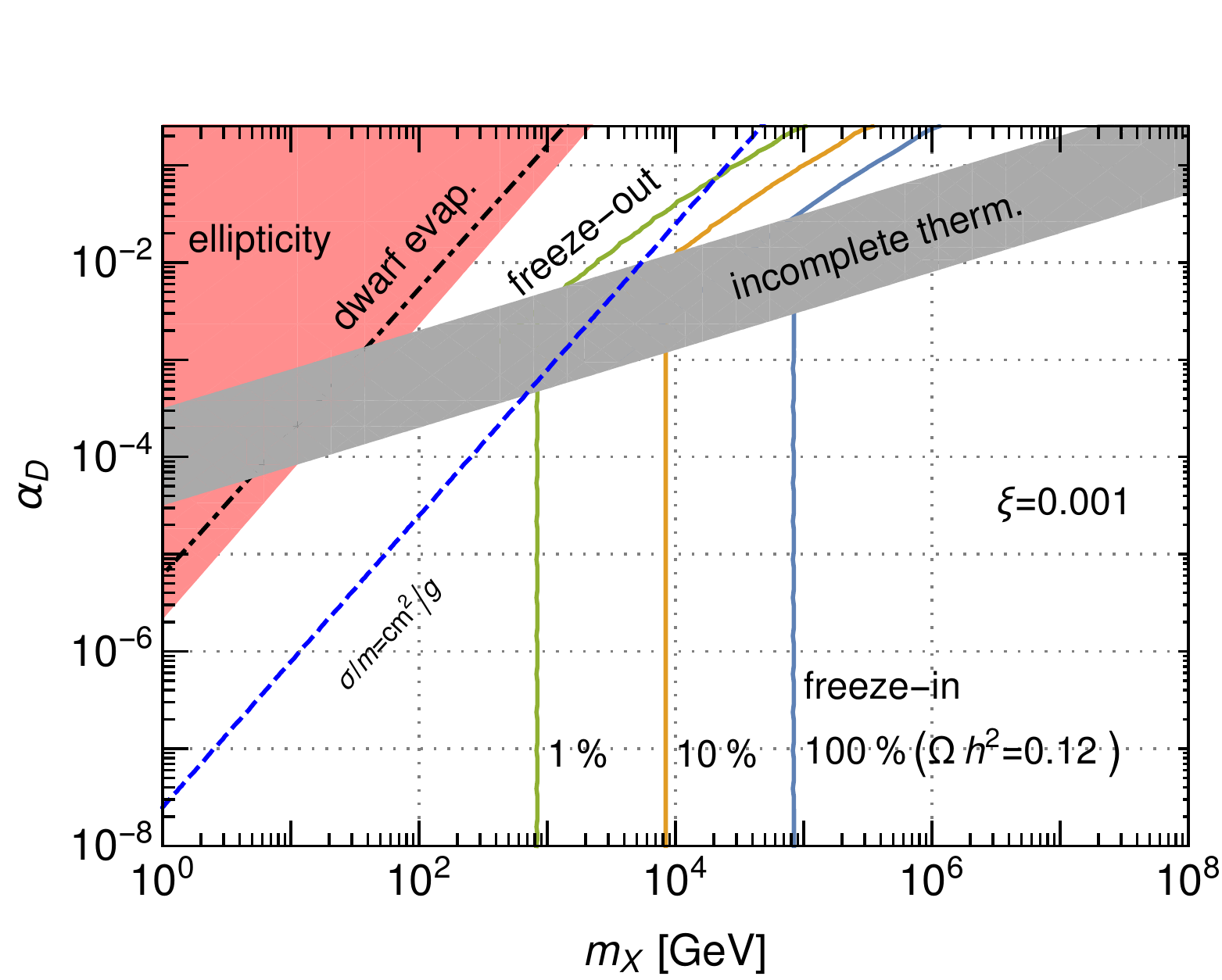}
\includegraphics[width=0.475\textwidth]{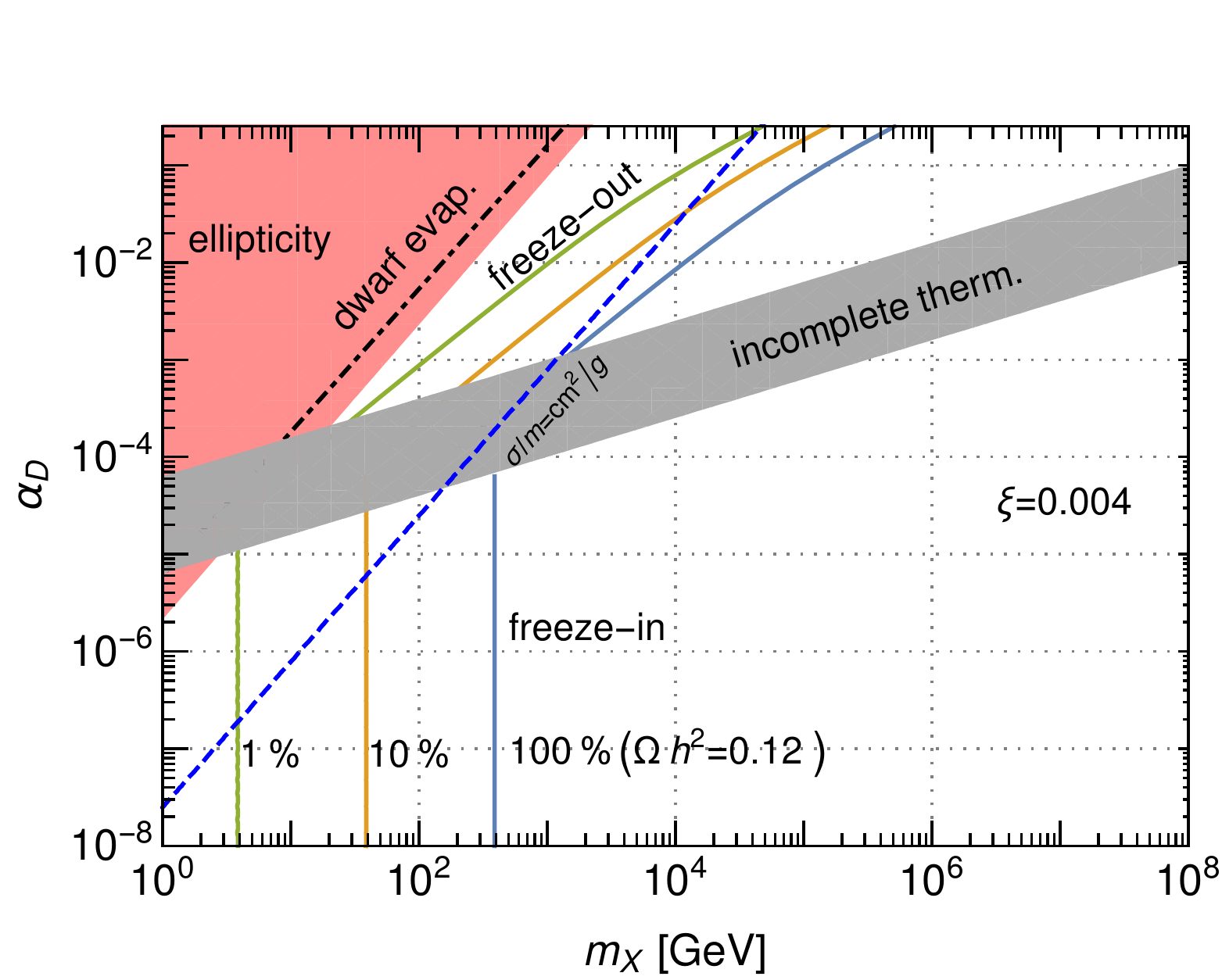}
\caption{Parameter space of the charged PIDM (mass versus dark fine-structure constant $\alpha_D$) for $\xi=0.001$ (corresponding to $T_{rh}=10^{-4}m_p$) and $\xi=0.004$ (corresponding to $T_{rh}=6\cdot 10^{-4}m_p$), respectively. 
The region above the grey shaded area corresponds to the
strong coupling regime $\alpha_D\geq \alpha_D^{\rm crit}$, and below to the weak coupling regime, $\alpha_D\leq \alpha_D^{\rm crit}/10$. In each region
we show contour lines of the PIDM abundance $\Omega_Xh^2=0.12,0.012,0.0012$, corresponding to $(100,10,1)\%$ of the measured DM density.
The region above the black dot-dashed line is excluded from observations of dwarf galaxy evaporation, and the red region from the ellipticity of the gravitational potential of NGC720 \cite{Agrawal:2016quu}. The blue dashed line corresponds to $\sigma/m=1{\rm cm}^2/{\rm g}$ for $v=30$km$/$s. }
\label{plot1}
\end{centering}
\end{figure}

\subsubsection{Strong coupling regime  $\alpha_D\gg \alpha_D^{\rm crit}$}

In this portion of parameter space, the dark sector is in thermal equilibrium at the beginning of the non-relativistic regime
with temperature $T_D=\xi T_{SM}$. In addition, for $a_{nr}<a<a_{\rm kin}$, see Eq.\,\eqref{eq:akin}, Compton scattering keeps the PIDM
temperature equal to the dark gauge boson temperature, i.e. $T_X=T_{\gamma_D}\equiv T_D\propto 1/a$.
We find that the portion of parameter space for which freeze-out is relevant ($\alpha\gg \alpha_D^{\rm crit}$), and for viable
values of $T_{rh}$, freeze-out of $X\bar X\to \gamma_D\gamma_D$ occurs \emph{before} kinetic decoupling for parameters that
are consistent with the overclosure constraint $\Omega_Xh^2\leq 0.12$ (see below).
The setup is therefore analogous to conventional thermal freeze-out, except for the different temperature in the dark sector, $\xi=T_D/T_{SM}$.

For $T\lesssim m_X$ the equilibrium number density is given by 
\be
n_X^{eq}(T) = \frac{g_X T m_X^2}{2\pi^2}K_2(m_X/T) \simeq g_X \left(\frac{Tm_X}{2\pi}\right)^{3/2}e^{-m_X/T}\,,
\ee
where $K_2$ is a modified Bessel function. The annihilation rate can be estimated as $\Gamma_{\rm ann}\simeq n_X^{eq}(T_D)\langle\sigma v\rangle$,
where $\langle\sigma v\rangle \simeq\pi\alpha_D^2\bar S_{\rm ann}(m_X/T_D)/m_X^2$ is the thermally averaged cross-section for $X\bar{X} \rightarrow \gamma_D \gamma_D$, including the Sommerfeld enhancement factor.
For $x\gtrsim 1$ the annihilation rate exceeds the Hubble rate for $\alpha\gg \alpha_D^{\rm crit}$ and $T_{rh}\ll 10^{-3}m_p$, such that PIDM annihilations remain
in equilibrium for some time in the non-relativistic regime, and $n_X\simeq n_X^{eq}$ decreases exponentially until $\Gamma_{\rm ann}$ drops below $H$. 

The resulting relic density has been estimated
within the freeze-out approximation in \cite{Agrawal:2016quu} (note that we use the convention $x=m_X/T_D$ involving the dark sector temperature),
\beq\label{omega}
\Omega_X\simeq \frac{16 \pi^3}{9 \sqrt{5 \pi}} \frac{g_{0S}}{\sqrt{g_{\rm eff}}} \frac{T_0^3}{m_{p}^3} \frac{\xi x_f (1+n)}{\langle\sigma v\rangle|_{x_f} H_0^2}, 
\eeq
where $T_0$ and $H_0$ are the CMB temperature and the Hubble parameter today, $g_{0S}=3.91$ and $g_{\rm eff}$ the effective number of SM degrees of freedom today and at freeze-out, respectively, $n=-d\ln\langle\sigma v\rangle/d\ln a|_{x_f}$ and $x_f$ is given as a solution to the equation
\beq
\xi \frac{\sqrt{90}}{8 \pi^2} \frac{g_X}{\sqrt{g_{\rm eff}}} \frac{\alpha_D^2m_{p}}{m_X} \delta (\delta+2)=\omega=\frac{\sqrt{ x_f}e^{x_f}}{\bar S_{\rm ann}(x_f)}, 
\eeq
where $\delta(\delta+2)\approx n+1$ is matched to the numerical freeze-out computation \cite{Kolb:1990vq} and $g_X=g_{\bar X}=2$ is the number of $X$ degrees of freedom. 
An approximate solution for $x_f$ is
\beq\label{xf}
x_f\simeq \ln\omega - \frac{1}{2} \ln( \ln\omega) + \ln(\bar S_{\rm ann}(\log\omega)).
\eeq
In order to assess whether the condition $a_f<a_{\rm kin}$ for kinetic equilibrium is satisfied, where $a_f=x_fa_{nr}$ is the scale-factor at freeze-out,
it is sufficient to obtain a rough estimate of $x_f$. Combining \eqref{omega} and \eqref{xf}, one can estimate that the freeze-out value $x_f$ required to
obtain the measured dark matter density is modified compared to the conventional freeze-out as $x_f \sim 25 +2\ln(\xi)$. Imposing the conservative
condition $T_{rh}/m_p<10^{-3}$ implies $x_f \lesssim 15$. On the other hand, the condition  $a_f<a_{\rm kin}$ requires
\be
  x_f < \frac{a_{\rm kin}}{a_{nr}} \simeq 3\cdot 10^3\,\left(\frac{\alpha_D}{10^{-2}}\right)\left(\frac{100\,{\rm GeV}}{m_X}\right)^{1/2}\left(\frac{T_{rh}}{10^{-4}m_p}\right)^{3/4}\,.
\ee
This condition is safely satisfied for the benchmark scenario we are mostly interested in, but could become violated for very large $m_X$ or small $T_{rh}$ or $\alpha_D$.
However, it turns out that the condition is always satisfied in the portion of parameter space that corresponds to the strong coupling regime, for which
freeze-out is relevant.

The authors of \cite{Agrawal:2016quu} consider $\xi=0.5$. For this value of $\xi$ the correct relic density is obtained for $\alpha_D \sim 0.001$ and $m_X \sim 100 \,\si{GeV}$. In the strongly coupled PIDM scenario $\xi=(T_{rh}/m_p)^{3/4}\ll 10^{-2}$ and $\alpha_D\gg \alpha_D^{\rm crit}$, which means that, keeping $m_X$ fixed, the relic density is approximately a factor of $(10^3\alpha_D)^2/\xi$ smaller (see Eqs. \eqref{omega} and \eqref{xf}). Therefore, freeze-out can yield the measured DM abundance only for much
larger $m_X\gtrsim 10^4$\,GeV (see below).
For our numerical results, we computed the relic density by numerically solving the Boltzmann equation, taking the thermally averaged Sommerfeld enhancement factor
into account.

The abundance of dark gauge bosons within the strongly coupled regime is given by $\rho_{\gamma_D}=\frac{\pi^2}{15}T_D^4$.
Due to the freeze-out in the dark sector, the abundance is slightly increased. We estimate this effect by assuming that the entropy
density is separately conserved within the dark sector, i.e. $s_D\propto a^{-3}$. For $T_D\gg m_X$ it is given by $s_D=\frac{\pi^2}{45}\left(g_{\gamma_D}+\frac78 (g_X+g_{\bar X})\right)T_D^3$, and $s_D\simeq \frac{\pi^2}{45}g_{\gamma_D}T_D^3$ for $T_D\ll m_X$. Entropy conservation thus implies $T_D(<m_X)/T_D(>m_X)\simeq (11/4)^{1/3}$.
This is similar to the annihilation of $e^+e^-$ pairs in the SM, which increases the photon temperature by the same factor.
Therefore, in the strongly coupled regime, using \eqref{xifulleq} we obtain for $T_D\ll m_X$
\be\label{eq:xiloweq}
  \xi = \frac{T_D}{T_{SM}} = (11/4)^{1/3}\left(\frac{T_{rh}}{m_p}\right)^{3/4}\,\left(\frac{g_*(T_{SM})}{g_{rh}}\right)^{1/3}\,.
\ee
At low temperatures $T_{SM}\ll$ MeV, this implies
\beq\label{eq:Neff2}
  \Delta N_{\rm eff} \simeq 0.2\left(\frac{T_{rh}}{m_p}\right)^3\,.
\eeq

\subsubsection{Result}

We show the result for the PIDM abundance $\Omega_X h^2$ within the weak- and strong coupling regimes in Fig.\,\ref{plot1}
for $T_{rh}=10^{-4}m_p$ and the maximal value $T_{rh}=6\cdot 10^{-4}m_p$, corresponding to a temperature ratio within
the strong coupling regime (and for the temperature regime
$T_{SM}\gg 100$GeV, $T_D\gg m_X$) of $\xi=T_{D}/T_{SM}\simeq (T_{rh}/m_p)^{3/4}\simeq 0.001$ 
and $\xi\simeq0.004$, respectively\footnote{Today, $\xi$ is smaller by a factor $(11/4)^{1/3}(g_{0S}/g_{rh})^{1/3}\simeq 0.5$.}. 
Above the grey shaded area, the dark sector thermalizes to a temperature $T_D=\xi T_{SM}$ and the PIDM abundance is set by freeze-out
in the dark sector. Nevertheless, the preceding freeze-in also plays a role, setting the value of $\xi$. The contour lines show which
combination of PIDM mass and coupling will yield $100\%$, $10\%$ or $1\%$ of the measured DM density. For smaller
values of $\alpha_D$, below the grey area, the PIDM abundance is solely determined by freeze-in, and the dark sector never
thermalizes. The corresponding contours of $\Omega_X h^2/0.12$ are shown as well. Within the grey area, thermalization is incomplete once the PIDM
turns non-relativistic. While we do not attempt to model this transition region here, we emphasize that it is a viable region of parameter space,
and expect the relic density contours to smoothly connect both regions. It is interesting to note that, in the strongly coupled regime, the
final DM abundance is rather insensitive to the reheating temperature, approximately $\Omega_X h^2 \propto \xi\propto (T_{rh}/m_p)^{0.75}$, while $\Omega_X h^2 \propto (T_{rh}/m_p)^{3}$ in the weakly coupled regime. Thus, the freeze-out $X\bar X\to \gamma_D\gamma_D$ within the dark sector, occurring after the initial 
freeze-in production of dark sector particles, effectively cancels the relatively strong dependence on $T_{rh}$ of the number densities of $X$ and $\gamma_D$
obtained from gravitational production.

For illustration, we also include constraints on the parameter space taken from Ref.\,\cite{Agrawal:2016quu}, related to the evaporation of dwarf galaxies
as well as the reduction of galactic ellipticities due to strong self-interactions. In addition, we indicate for which values of parameters $\sigma/m=1{\rm cm}^2/{\rm g}$ 
for $v=30$km$/$s.

Note that both in the weakly and strongly coupled regime, the abundance of dark gauge bosons gives a negligible contribution to $\Delta N_{\rm eff}$,
see Eqs.\,\eqref{eq:Neff} and \eqref{eq:Neff2}, respectively, such that constraints on extra radiation are safely satisfied. As mentioned before, this property constitutes a testable difference to the scenario of hidden charged dark
matter.

\subsection{GUT scale charged PIDM}\label{GUT}

So far, we focused on the regime in which $m_X\ll T_{rh}$, which is relevant when insisting on a sizeable self-interaction
cross section  Eq.\,\eqref{sigma/m}.
Here we turn to the minimal PIDM scenario, for which the dark matter mass is close to the GUT scale, $m_X \sim 10^{-3} m_p$, and the correct relic abundance is obtained for instantaneous reheating with a temperature of $T_{rh} \sim 10^{-4} m_p$. Clearly, a dark matter particle this heavy cannot resolve the discrepancies between numerical simulations and observations on galactic scales, since the self-interaction cross section is strongly suppressed. However, if SIDM is not responsible for resolving these issues, we may entertain the idea of GUT scale charged dark matter. 

In this scenario, when PIDM particles are produced by the SM plasma, they are already non-relativistic. Given that the PIDM is so heavy, we expect dark photons to vastly dominate in number after freeze-in production is complete. This is indeed the case, as one can see by estimating the final number density from Eq.\,\eqref{niX} in the two opposite mass limits. Dark photons are produced with the same number density as computed previously, $n_{i,\gamma_D}\simeq 0.65 T_{rh}^6/m_p^3$. 

Since the PIDM is non-relativistic, the corresponding cross section is affected by Sommerfeld enhancement, similar as for freeze-out.
This can be taken into account by multiplying the right-hand side of Eq.\,\eqref{eq:sigXX} by $\bar S_{\rm ann}(m_X/T)$,   given by Eq.\,\eqref{avsomm}.
In the limit $T\ll m_X$ this gives
\be
 \langle\sigma v\rangle_{X\bar X \to {\rm SM}\,{\rm SM}'} = \frac{230\pi m_X T}{m_p^4}\bar S_{\rm ann}(m_X/T)\,.
\ee
Sommerfeld enhancement can play a role if $\pi\alpha_D\gtrsim \sqrt{T_{rh}/m_X}$.
Otherwise, $\bar S_{\rm ann}$ is of order unity, which we assume for the estimate below.
Using Eq.\,\eqref{eq:abundance} and assuming $T_{rh}\ll m_X$, we obtain 
\be
n_{i,X}\simeq \frac{115\kappa_2^2}{4\pi^2}\frac{m_X^3T_{rh}^3}{m_p^3} \exp(-2m_X/T_{rh})
\simeq 0.18\frac{m_X^3T_{rh}^3}{m_p^3} \exp(-2m_X/T_{rh})\,, 
\ee
so that the PIDM number density is exponentially suppressed with respect to the dark photon number density. In particular we find 
$n_{\gamma_D}/n_X\simeq 3(T_{rh}/m_X)^3\exp(2m_X/T_{rh})$.
For the benchmark values quoted above this gives $n_{\gamma_D}/n_{X} \sim 10^{6}$. 

When the PIDM is produced it is already non-relativistic, and freeze-out does not occur in this scenario. However, since the two number densities are so different, it is in principle possible for the dark photons to pair create dark matter particles and increase the relic abundance. In order to estimate whether this
may affect the PIDM density, we consider the contribution to the Boltzmann equation for freeze-in from pair creation,
and define a corresponding rate by $\Gamma \equiv \frac{d}{dt}\ln(a^3n_X)_{\gamma \gamma \rightarrow X \bar{X}}$.
It is given by
\bea
\Gamma_{\gamma \gamma \rightarrow X \bar{X}} &=& \frac{1}{n_X} n_{\gamma_D}^2 \langle\sigma v\rangle _{\gamma \gamma \rightarrow X \bar{X}}
  =\frac{1}{n_X} \left(\frac{n_{\gamma_D}}{n_{\gamma_D}^{eq}}\right)^2(n_X^{eq})^2\langle\sigma v\rangle _{X \bar{X}\to\gamma_D\gamma_D}\nn\\
  &\simeq& \frac{1}{n_X} \left(\frac{n_{\gamma_D}}{n_{\gamma_D}^{eq}}\right)^2(n_X^{eq})^2\frac{\pi \alpha_D^2}{m_X^2} \bar{S}_{\rm ann}(m_X/T)\,,
\eea
where we approximate the average over the distribution
function by a thermal average with temperature of order $T_{SM}$. In particular, the equilibrium densities are evaluated for $T_{SM}$.
We compare this rate to the corresponding rate for gravitational production, 
\be
  \Gamma_{grav}=\frac{1}{n_X} (n_X^{eq})^2\langle\sigma v\rangle _{X \bar{X}\to{\rm SM}{\rm SM}'}\simeq \frac{1}{n_X} (n_X^{eq})^2\frac{230\pi m_X T}{m_p^4}\bar S_{\rm ann}(m_X/T)\,.
\ee
Their ratio is for $T_{rh}\ll m_X$ given by 
\be
  \frac{\Gamma_{\gamma \gamma \rightarrow X \bar{X}}}{\Gamma_{grav}}
  \simeq \left(\frac{n_{\gamma_D}}{n_{\gamma_D}^{eq}}\right)^2\frac{\alpha_D^2m_p^4}{230m_X^3T}
  \simeq 0.06\frac{\alpha_D^2T_{rh}^5}{m_X^3m_p^2}\ll 1\,,
\ee
where we used that the dominant contribution comes from $T\simeq T_{rh}$ in the last step, and inserted
the dark gauge boson density using Eq.\,\eqref{niX}.
Therefore, even though the cross section for $\gamma \gamma \rightarrow X \bar{X}$ is enhanced by a factor of order $m_p^4/(m_X^3T)$
compared to gravitational production, this channel is suppressed because $(n_{\gamma_D}/n_{\gamma_D}^{eq})^2\sim (T_{rh}/m_p)^6\ll 1$.

Therefore, the $U(1)$ interaction can only affect the production of GUT scale PIDM via Sommerfeld enhancement of the gravitational production,
for very large values of the fine-structure constant $\pi\alpha_D\gg \sqrt{T_{rh}/m_X}$. 
The contribution to the extra radiation density parameterized by $\Delta N_{\rm eff}$ is given by the same expressions Eq.\,\eqref{eq:Neff}
as in the case of a low mass PIDM, and therefore also strongly suppressed.

\section{Particle physics models for $\xi = O(1)$}\label{xiO1}

In section \ref{pheno} we showed that if the dark sector is maximally decoupled from the visible sector, then 
$\Delta N_{\rm eff}\leq 4\cdot 10^{-11}$, obtained from combining the result from gravitational production
of dark gauge bosons, Eq.\,\eqref{eq:Neff2}, which implies the scaling $\Delta N_{\rm eff} \propto (T_{rh}/m_p)^3$,
with the maximal value of the reheating scale given the bound $r\leq 0.064$ on the tensor-to-scalar ratio.
If the dark sector thermalizes, this also implies $\xi=T_D/T_{DM}\simeq (11/4)^{1/3}(T_{rh}/m_p)^{3/4}(g_{*}(T_{SM})/g_{rh})^{1/3}\leq 0.004$.

These are generic predictions due to the gravitational coupling between the dark and visible sector.
Any additional coupling will tend to increase $\Delta N_{\rm eff}$ and $\xi$.
The hidden charged dark matter scenario~\cite{Agrawal:2016quu} with $\xi =0.5$ therefore
requires a stronger-than-gravitational interaction between the two sectors.
Suppose for example that the two sectors communicate through a massive mediator $B$ with mass $m_B$, and renormalizable coupling $g_{BX\bar X}$ to the
dark sector as well as $g_{B{\rm SM}{\rm SM}'}$ to a pair of SM particles. The dark matter particles and the thermal bath of SM particles created after reheating
may establish thermal equilibrium between the two sectors via $B$ exchange. The temperatures of the two sectors are then equal to each other, $T_{SM}=T_D \equiv T$. If the thermally averaged cross section $\langle\sigma v\rangle _B$ for the process $X+\bar{X} \leftrightarrow SM+SM$ mediated by $B$ is not suppressed by any scale larger than the masses, then for $T\gg m_B, m_X$ it generically scales as $\langle\sigma v\rangle _B \sim \pi\alpha_B^2 T^{-2}$, where $\alpha_B\equiv g_{BX\bar X}g_{B{\rm SM}{\rm SM}'}/(16\pi^2)$. The interaction rate $\Gamma_B = n_X \langle\sigma v\rangle _B \simeq 2\zeta(3)\alpha_B^2 T/\pi$ exceeds the Hubble rate $H \simeq T^2/(\kappa_2^2m_p)$
for $\alpha_B^2 \gg \pi T/(2\zeta(3)\kappa_2^2m_p)\simeq 20\,T/m_p$. Once $T$ drops below $m_B$ or $m_X$, the rate $\Gamma_B$ becomes suppressed.
Therefore, for $T_{rh} \gg m_B \gg m_X$, $B$-exchange can establish thermal equilibrium between the dark and visible sector if
\be
m_B \ll \frac{\alpha_B^2}{20} m_p \qquad ({\rm for}\ T_{rh} \gg m_B)\,.
\ee
The two sectors remain in thermal equilibrium until $B$ becomes nonrelativistic, at which point the cross section is suppressed by a factor of $m_B^{-4}$, the Hubble rate quickly overcomes the interaction rate and the dark sector decouples from the SM plasma. 

On the other hand, if $m_B \gg T_{rh} \gg m_X$, then $\langle\sigma v\rangle _B \sim \pi\alpha_B^2 T^2/m_B^4$  and the condition to maintain thermal equilibrium gives 
$T \ll m_B \ll 0.5\sqrt{\alpha_B} T^{3/4}m_p^{1/4}$, which can be satisfied in a narrow range of temperatures if $T_{rh}$ is sufficiently small. 
In particular,
\be
m_B \ll 0.5\sqrt{\alpha_B} T_{rh}^{3/4}m_p^{1/4} \qquad ({\rm for}\ T_{rh} \ll m_B)\,.  
\ee
Now we want to use this discussion to estimate the maximal mass $m_B$ such that thermal equilibrium is reached. This depends on the value of the gauge coupling $\alpha_B$. If $\alpha_B^2/20 > T_{rh}/m_p$, the maximal value is obtained in the second case, i.e. $m_B^{\rm max}=0.5\sqrt{\alpha_B} T_{rh}^{3/4}m_p^{1/4}>T_{rh}$.
If $\alpha_B^2/20 < T_{rh}/m_p$, the thermal equilibrium cannot be reached in the case $T_{rh} \ll m_B$, and therefore $m_B^{\rm max}=\alpha_B^2m_p/20 < T_{rh}$.

\begin{figure}[th]
\centering
\includegraphics[width=12cm]{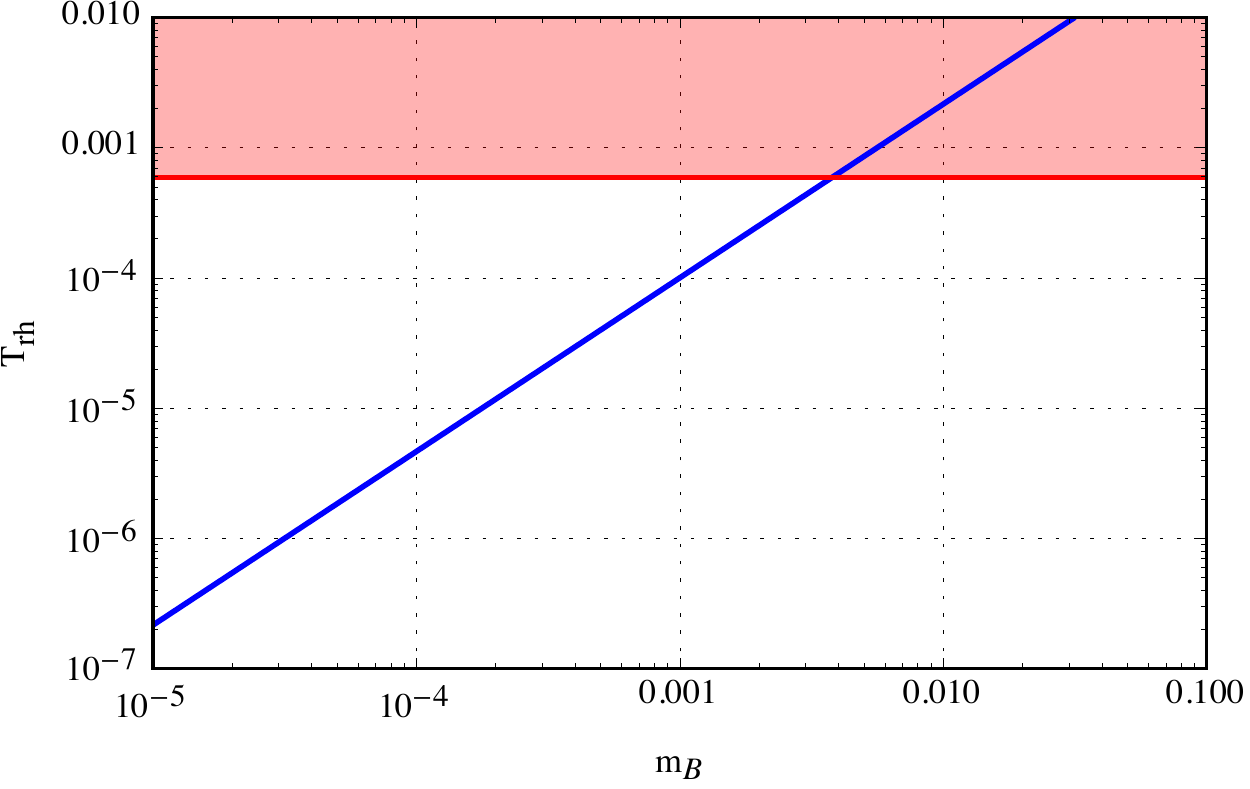}
\caption{\label{plot2} Constraint on the mediator mass $m_B = \sqrt{\alpha_B} (T_{rh}^3 m_p)^{1/4}$ in blue, 
with the assumption that the coupling constant $\alpha_B$ is order one. 
The region above the red line is excluded by the upper bound $r\leq 0.064$ on the tensor-to-scalar ratio. The parameter space above the blue line and not excluded by the red line can give rise to thermal equilibrium between the dark and visible sectors. Masses above the GUT scale (0.01 $m_p$) are excluded. All quantities are given in Planck units.}
\end{figure}

Assuming $T_{rh}\lesssim 10^{-4}m_p$, the largest possible mass compatible with perturbative couplings is 
\be
  m_B^{\rm max}\simeq T_{rh}^{3/4}m_p^{1/4}\lesssim 10^{-3}m_p\,,
\ee 
i.e. around the GUT scale. If the mass is higher than the GUT scale, thermal equilibrium will never be reached.
The constraint on the mediator mass is illustrated in Fig.\,\ref{plot2}. 

If thermal equilibrium is reached, the temperature ratio today is given by
\beq
\xi = \frac{T_D}{T_{SM}} = \left(\frac{g_{0S}}{g_i}\right)^{1/3},
\eeq
where $g_{0S}\simeq 3.91$ and $g_i$ are the number of relativistic degrees of freedom today and when the dark and
visible sectors decouple, with $g_i\simeq 106.75$ for $T_{SM}\gg 100$\,GeV, giving $\xi \simeq 0.33$ and $\Delta N_{\rm eff}\simeq 0.054$.

One could also view the mediator mass $m_B$ as the energy scale that suppresses the non-renormalizable operator which describes the interaction at low energies, i.e. the Fermi interaction $(\alpha_B/m_B^2) (\bar{\psi}_{SM} \gamma^\mu \psi_{SM})(\bar{\psi}_X \gamma_\mu \psi_X)$ (for a spin-$1$ mediator $B$ with vector couplings). 
This description is applicable in the high mass regime $m_B \gg T_{rh}$.
If $m_B\gg 0.5\sqrt{\alpha_B} T_{rh}^{3/4}m_p^{1/4}$, thermal equilibrium is not reached. 
Instead,  dark sector particles are produced via freeze-in.
This case can be treated  analogously to production of dark sector particles via gravitational interaction, with suppression
scale of the cross section given by $(m_B/\sqrt{\alpha_B})^{-4}$ instead of $m_p^{-4}$.
If the dark sector thermalizes within itself, the resulting temperature ratio is of order
\be
  \xi \sim \frac{\sqrt{\alpha_B}}{m_B} \left(T_{rh}^3m_p\right)^{1/4}\left(\frac{g_{*}(T_{SM})}{g_{rh}}\right)^{1/3}\,.
\ee
The ratio becomes of order one,  corresponding to thermal equilibrium, for $m_B \sim \sqrt{\alpha_B} (T_{rh}^3 m_p)^{1/4}$.
This coincides with the condition that we found previously. In this case, we find
\be
\Delta N_{\rm eff} \sim 0.01 \left(\frac{\alpha_B}{0.3}\right)^2 \left(\frac{10^{-16}\,{\rm GeV}}{m_B}\right)^4\left(\frac{T_{rh}}{10^{-4}m_p}\right)^3\,.
\ee
As discussed in Sec.\,\ref{sec:abundance}, the right-hand side is enhanced by a factor $(11/4)^{4/3}\simeq 4$ if the dark
sector is in thermal equilibrium within itself while the PIDM becomes non-relativistic.

As mentioned before, future CMB and large-scale structure observations~\cite{Baumann:2015rya, Baumann:2017gkg} 
could discriminate between the thermalized scenario with $\Delta N_{\rm eff}\simeq 0.054$
and the case where the dark- and visible sectors have never been in thermal equilibrium, giving $\Delta N_{\rm eff}\lesssim 0.054$, saturating
the inequality for $m_B \sim \sqrt{\alpha_B} (T_{rh}^3 m_p)^{1/4}$.

\section{Conclusions}\label{conclusions}

We have considered the scenario of charged PIDM, where dark matter is maximally hidden from the SM, but charged under its own unbroken $U(1)$ gauge symmetry. With only gravitational interactions between the PIDM and the SM, the interactions are too weak to establish thermal equilibrium of the PIDM with the SM thermal bath.
However, they may establish equilibrium separately within the dark sector between the PIDM and the dark photons for a dark fine-structure constant above a critical value
 of order $\alpha_D \sim 10^{-3}$ (see Eq.\,\eqref{eq:alphacrit}). We have provided a qualitative description as well as parametric estimates of the thermalization dynamics to arrive at the critical value. The measured DM abundance can be obtained for $\alpha_D$ above and below the critical value. In the latter case, $\Omega_X\propto T_{rh}^3$ has a relatively strong dependence on the reheating temperature, due to freeze-in production via gravitational interactions. If $\alpha_D \gtrsim 10^{-3}$, the PIDM abundance is further modified by freeze-out within the dark sector. Interestingly, even though the initial PIDM population is still produced via gravitational freeze-in, the subsequent freeze-out dynamics largely compensates the dependence on the reheating temperature, giving $\Omega_X\propto T_{rh}^{0.75}$.

The measured DM abundance can be produced for $m_X \gtrsim 100$GeV and over a wide range of values for $\alpha_D$. This covers a potentially interesting parameter range of the dark matter mass and dark fine-structure constant for the small-scale issues in structure formation according to the analysis \cite{Agrawal:2016quu}. This motivates further studies on the extent to which an unbroken gauge symmetry in the dark sector is indeed a viable option.

The charged PIDM has a different thermal history than assumed in previously proposed models of ``hidden charged dark matter'', which assumed stronger than gravitational interactions with the SM and initial thermal equilibrium with the SM bath \cite{Foot:2004pa,Feng:2008mu,Ackerman:mha,Feng:2009mn,Feng:2009hw,Das:2010ts,Agrawal:2016quu}. These models can be discriminated due to different predictions for the effective number of neutrino species, $N_{\rm eff}$. In addition, as in the uncharged case \cite{Garny:2015sjg,Garny:2017kha}, the charged PIDM also requires observable primordial tensor modes, if the mass is in the $m_X \sim 100$  GeV regime, in order for dark matter to be produced with the right abundance in the early universe.

\bigskip

{\bf \noindent Acknowledgements}

\smallskip

\noindent We thank Patrick Vaudrevange for helpful discussions. AP and MSS are supported by Villum Fonden grant 13384. CP3-Origins is partially funded by the Danish National Research Foundation, grant number DNRF90. MSS would also like to thank the CERN theoretical physics department for kind hospitality while parts of this work was completed.

\begin{appendix}
\numberwithin{equation}{section}

\setcounter{equation}{0}

\section{Averaged annihilation cross section for non-thermal distributions}\label{appendix}

In this appendix we discuss the dependence of the averaged annihilation cross section
for the process $X\bar X\to \gamma_D\gamma_D$ on the shape of the distribution function.
As an example, we compare the conventional thermal average, obtained for a Maxwell-Boltzmann distribution
function $f_{MB}(p) \propto e^{-E_p/T}$, where $E_p=\sqrt{m^2+p^2}$, with the (somewhat extreme)
case of a narrowly peaked distribution, given by $f_D(p) \propto \delta(E_p-\bar E)$.
We require that both distributions correspond to an average energy of the same
order of magnitude, which can be achieved by setting $\bar E=3T$ in the relativistic
regime and $\bar E=m+\frac32 T$ in the non-relativistic regime.

The averaged annihilation cross section is defined as \cite{Gondolo:1990dk}
\be\label{sigmav}
\langle\sigma v\rangle = \frac{\int d^3p_1 d^3p_2\,\sigma v_{M\o l} f(p_1) f(p_2) }{\int d^3p_1d^3p_2\,f(p_1)f(p_2)},
\ee
where $p_1$ and $p_2$ are the momenta of the colliding particles, $f(p_1)$ and $f(p_2)$ their (generic) distribution functions, and $v_{M \o l}$ the M\o ller velocity. In our case the incoming particles are massive with mass $m\equiv m_X$ and the outgoing particles (dark photons $\gamma_D$) are massless. For Maxwell-Boltzmann distribution functions, the thermal average can be expressed as \cite{Gondolo:1990dk}
\be\label{gondologelmini}
\langle\sigma v\rangle^{MB}=\frac{1}{8 m^4 T K_2^2(m/T)}\int_{4m^2}^{\infty}ds\,\sigma (s-4m^2)\sqrt{s}K_1(\sqrt{s}/T)\,,
\ee
where $K_1(x)$ and $K_2(x)$ are the modified Bessel functions of the second kind and $s$ is the center of mass energy squared. 

At tree-level, the total cross section $\sigma(s)$ for $X\bar{X}\rightarrow \gamma_D \gamma_D$ annihilation is given by
\beq\label{sigmaann}
\sigma(s)=\frac{2 \pi  \alpha_D ^2 \left(\left(-8 m^4+4 m^2 s+s^2\right) \log \left(\frac{\sqrt{s-4 m^2}+\sqrt{s}}{\sqrt{s}-\sqrt{s-4 m^2}}\right)-\sqrt{s \left(s-4 m^2\right)} \left(4 m^2+s\right)\right)}{s^2 \left(s-4 m^2\right)}.   
\eeq
Inserting the total cross section \eqref{sigmaann} into \eqref{gondologelmini} and taking the high energy limit $\sqrt{s} \gg m$, one obtains 
\beq\label{svrel}
\langle\sigma v\rangle_{rel}^{MB} = \frac{2 \pi  \alpha_D ^2 T^2 (\log (2 T/m)-\gamma_E )}{m^4 K_2\left(\frac{m}{T}\right){}^2} \to \frac{\pi  \alpha_D ^2}{2T^2} \left( \log \left(\frac{2 T}{m}\right)-\gamma_E\right)\,,
\eeq 
where $\gamma_E$ is the Euler-Mascheroni constant. This expression is further simplified by taking the ultra-relativistic limit $m \rightarrow 0$ in the last step.

Taking instead the non-relativistic limit, we obtain for the thermally averaged cross section 
\beq\label{svnonrel}
\langle\sigma v\rangle_{non-rel}^{MB} = \frac{\pi ^2 \alpha_D ^2 T e^{-\frac{2 m}{T}} \left(4 m^2+6 m T+3 T^2\right)}{8 m^5 K_2\left(\frac{m}{T}\right){}^2} \to \frac{\pi \alpha_D^2}{m^2}\,,
\eeq
where the last step corresponds to the limit $T\to 0$, in which case only the constant $s$-wave contribution remains.
As we expect, the averaged cross section for annihilation of massive particles into dark photons scales like $\alpha_D^2/E^2$, where the typical energy of the process $E$ is of the order of the mass $m$ of the colliding particles in the non-relativistic limit and of the order of the temperature $T$ in the ultra-relativistic limit. 

In the non-relativistic regime, Sommerfeld enhancement has to be taken into account. In the $s$-wave limit, the tree-level cross section is multiplied
by the factor $S_{\rm ann}=\frac{2 \pi \alpha_D/v}{1-e^{-2 \pi \alpha_D/v}}$. Its average is given by (here we traded the momenta for velocities)
\beq\label{sann}
\bar{S}_{\rm ann} = \frac{\int  d^3\vec{v}_1d^3\vec{v}_2 S_{\rm ann}\,f(\vec{v}_1)f(\vec{v}_2) }{\int d^3\vec{v}_1d^3\vec{v}_2\,f(\vec{v}_1)f(\vec{v}_2) }\,.
\eeq
For a Maxwell-Boltzmann distribution, taken in the non-relativistic limit (see also Eq.\,\eqref{avsomm} in the main text),
one obtains  \cite{Feng:2010zp}
\beq\label{sannT}
\bar{S}_{\rm ann}^{MB}= \frac{x^{3/2}}{2 \sqrt{\pi}} \int_0^{\infty} S_{\rm ann}(\alpha_D/v) v^2 e^{-\frac{xv^2}{4}} dv\,,
\eeq
where $x=m/T$. The thermally averaged cross section in the non-relativistic limit is thus given by
\beq\label{svnonrelSE}
\langle\sigma v\rangle_{non-rel}^{MB} = \frac{\pi \alpha_D^2}{m^2} \times \bar{S}_{\rm ann}^{MB}\,.
\eeq

\medskip

Let us consider now the averaged cross section for the narrowly peaked distribution $f_D \propto \delta(E_p-\bar E)$.
The center of mass energy squared is $s=4m^2+2\bar p^2(1-\cos\theta)$, where $\bar p$  is the momentum
corresponding to $\bar E$, defined via $\bar E=\sqrt{m^2+\bar p^2}$, and $\theta$ the angle between the momenta of the
annihilating particles. The M\o ller velocity is
given by
\beq
v_{M \o l}=\frac{\sqrt{s(s-4m^2)}}{2 E_1 E_2}\big|_{E_i=\bar E}.
\eeq
The averaged cross section Eq.\,\eqref{sigmav} is 
\be
  \langle\sigma v\rangle^D = \frac12 \int_{-1}^1d\cos\theta\,\sigma v_{M \o l}=\frac{1}{8\bar p^2\bar E^2}\int_{4m^2}^{4\bar E^2}ds\,\sqrt{s(s-4m^2)}\sigma(s)\,.
\ee
In the relativistic limit $\sqrt{s} \gg m$ the total tree-level cross section Eq.\,\eqref{sigmaann}  becomes $\sigma(s) \simeq \frac{2 \pi \alpha_D^2}{s} (\log(s/m^2)-1)$ and $\bar E\simeq \bar p$. Putting everything together we get
\beq\label{svrelD}
\langle\sigma v\rangle_{rel}^D = \frac{2\pi \alpha_D^2}{\bar p^2} \left(\log \left(\frac{2\bar p}{m}\right)-1\right),
\eeq
which is of the same order of magnitude as the averaged cross section for a Maxwell-Boltzmann distribution with
the same average momentum ($\bar p\sim 3T$), see\,\eqref{svrel}.

\begin{figure}
\begin{center}
\includegraphics[width=0.6\textwidth]{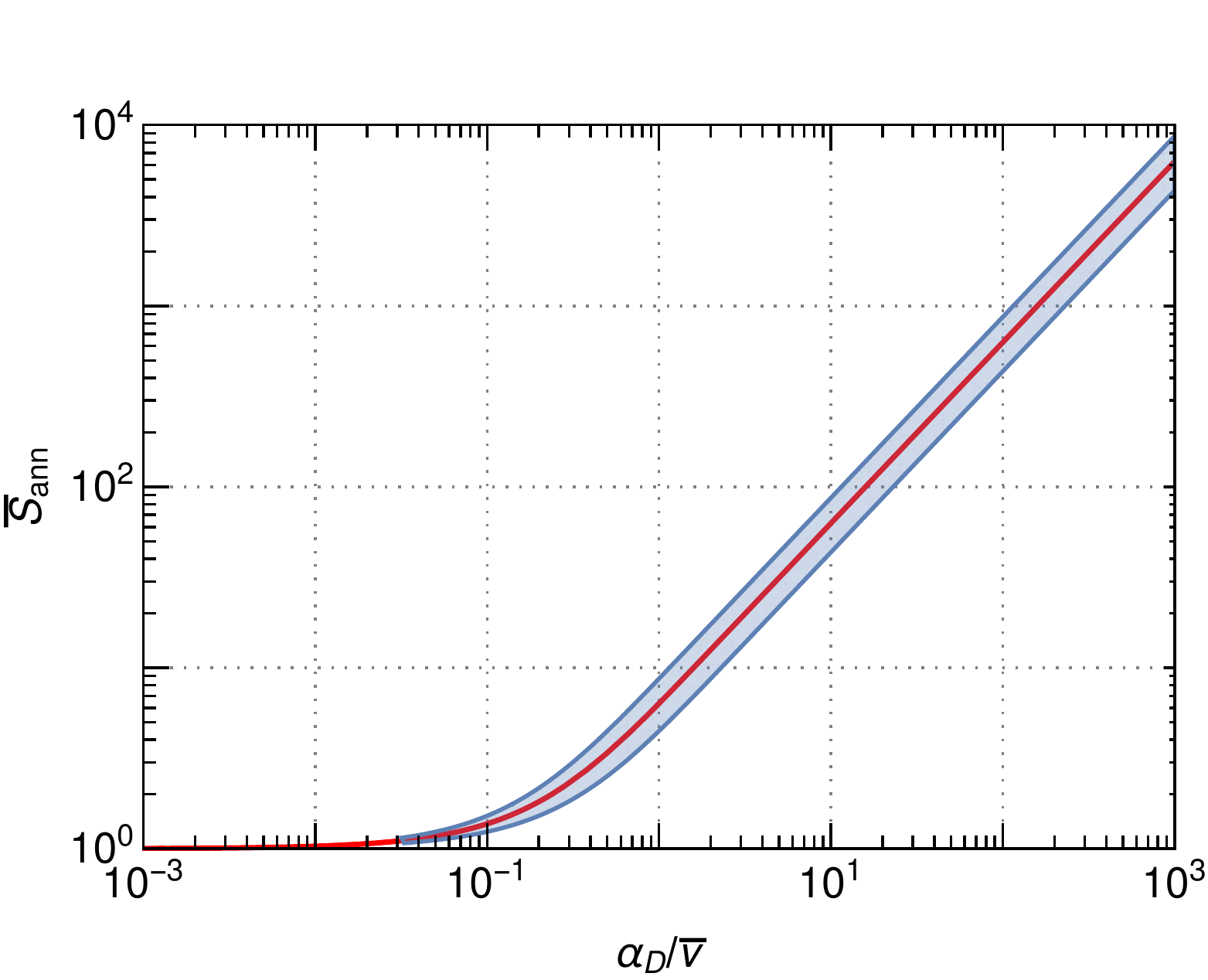}
\end{center}
\caption{\label{TvsD} Comparison of the averaged Sommerfeld enhancement factor computed for a narrowly peaked distribution function around 
an Energy $\bar E=m+\frac 12 m\bar v^2$ (red line)
and a thermal distribution function, with temperature $T$ adjusted such that the average energy $\langle E\rangle=m+\frac32 T$ is
of comparable size, more precisely $\frac12(\bar E-m)\leq\langle E\rangle-m\leq 2(\bar E-m)$ (shaded region). For $\langle E\rangle=\bar E$ the thermal result
is indistinguishable from the one for a peaked distribution on the scale of this figure, with $\bar{S}_{\rm ann}^{MB}/\bar{S}_{\rm ann}^{D}\to \sqrt{3/\pi}$
for $\alpha_D/\bar v\gg 1$ (see text for details).}
\end{figure}

For freeze-out within the dark sector, the non-relativistic regime is more relevant.
In the non-relativistic limit $\sigma(s)\simeq \pi \alpha_D^2/(m \sqrt{s-4m^2})$, and $\bar E\simeq m+\bar p^2/(2m)$. In this case
the averaged tree-level cross section
\beq\label{svnonrelD}
\langle\sigma v\rangle_{non-rel}^D= \frac{\pi \alpha_D^2}{m^2},
\eeq
agrees with the thermal case in the $s$-wave limit. This is expected because $\sigma v$ approaches a constant, and therefore the
integrals over distribution functions cancel in \eqref{sigmav}. For the averaged Sommerfeld enhancement factor we obtain
\beq\label{sannD}
\bar{S}_{\rm ann}^D= \frac12\int_{-1}^1d\cos\theta S_{\rm ann}(\alpha_D/v)=\frac{1}{2\bar v^2}\int_0^{2\bar v}dv\, v\,S_{\rm ann}(\alpha_D/v)\,,
\eeq
where $\bar v\equiv \bar p/m$ is the peak velocity and $v=\sqrt{2}\bar v\sqrt{1-\cos\theta}$ the relative velocity
in the non-relativistic limit. The averaged cross section in the non-relativistic limit is therefore
\beq\label{svnonrelSE_D}
\langle\sigma v\rangle_{non-rel}^{D} = \frac{\pi \alpha_D^2}{m^2} \times \bar{S}_{\rm ann}^{D}\,.
\eeq
The averaged Sommerfeld factor may be compared to the thermal case \eqref{sannT}
with comparable average energy, corresponding to $\bar E-m \simeq \frac12 m\bar v^2\sim \frac32 T$ or $\bar v\sim \sqrt{3/x}$ (see Fig.\,\ref{TvsD}).
For $\alpha_D/\bar v\ll 1$, the enhancement factor itself approaches unity and therefore $\bar{S}_{\rm ann}\to 1$ independent
of the shape of the distribution function. In the opposite limit $\alpha_D/\bar v\gg 1$, one finds 
$\bar{S}_{\rm ann}^{D}\to 2\pi\alpha_D/\bar v$, which corresponds to $S_{\rm ann}$ evaluated for $v=\bar v$.
For comparison, for a thermal distribution one can check that $\bar{S}_{\rm ann}^{MB}\to 2\alpha_D\sqrt{\pi x}=2\pi\alpha_D/\bar v \times \sqrt{3/\pi}$,
where we assumed $\bar v= \sqrt{3/x}$ in the last step. Therefore, the Sommerfeld enhancement factors are of comparable size.

We expect the statements from above to hold true qualitatively also for more general distribution functions, as long as they correspond to comparable
average energy or momentum as for a thermal distribution with given temperature $T$.

\end{appendix}

\end{document}